\documentclass[manuscript, screen]{acmart}

\usepackage{xspace}
\usepackage[breakable]{tcolorbox}
\usepackage{soul}
\usepackage{caption}
\usepackage{booktabs}
\usepackage{multirow}
\usepackage{tabularx}
\usepackage{lipsum} 
\usepackage{array}
\usepackage{subcaption}
\usepackage{graphicx}
\usepackage{adjustbox}

\AtBeginDocument{%
  }

\acmISBN{978-1-4503-XXXX-X/2018/06}

\newcommand{\toolname}{SemOpt\xspace}

\makeatletter
\def\code{%
  \begingroup
  \small\ttfamily
  \@activecode\@code
}
\def\@activecode{%
  \catcode`.\active \catcode`_\active \catcode`'\active \catcode``\active
}
\begingroup\@activecode
\gdef\@code#1{%
  \def.{\char46\allowbreak}%
  \def_{\_\allowbreak}%
  \let'\textquotesingle
  \let`\textasciigrave
  #1%
  \endgroup
}
\endgroup
\makeatother

\newtcolorbox{rqbox}{breakable,left=2pt,right=2pt,top=2pt,bottom=2pt}

\usepackage{xcolor}

\definecolor{revblue}{RGB}{0,80,200}

\usepackage{xparse} 
\newcommand{\mainpoint}[1]{%
  \par 
  \smallskip
  \noindent
  \textbf{#1}%
}

\usepackage{enumitem}
\setlist[itemize]{leftmargin=1.5em}
\setlist[enumerate]{leftmargin=1.5em}

\begin{document}

\title{SemOpt: LLM-Driven Code Optimization via Rule-Based Analysis}

\newcommand{\PKUinstitution}{Key Laboratory of High Confidence Software Technologies (Peking University), Ministry of Education; School of Computer Science, Peking University}

\author{Yuwei Zhao}
\affiliation{%
  \institution{\PKUinstitution}
  \city{Beijing}
  \country{China}}
\email{zhaoyuwei@stu.pku.edu.cn}

\author{Yuan-An Xiao}
\affiliation{%
  \institution{\PKUinstitution}
  \city{Beijing}
  \country{China}}
\email{xiaoyuanan@pku.edu.cn}

\author{Qianyu Xiao}
\affiliation{%
  \institution{\PKUinstitution}
  \city{Beijing}
  \country{China}}
\email{2200012932@stu.pku.edu.cn}

\author{Zhao Zhang}
\affiliation{%
  \institution{\PKUinstitution}
  \city{Beijing}
  \country{China}}
\email{zhangzhao2019@pku.edu.cn}

\author{Yingfei Xiong}
\authornote{Corresponding author}
\affiliation{%
  \institution{\PKUinstitution}
  \city{Beijing}
  \country{China}}
\email{xiongyf@pku.edu.cn}


\renewcommand{\shortauthors}{Zhao et al.}

\begin{abstract}

Automated code optimization improves program performance through refactoring, and recent studies leverage LLMs for this purpose. Existing approaches mine optimization commits from open-source codebases to build large-scale knowledge bases, then employ retrieval techniques such as BM25 to obtain relevant examples for hotspot code, guiding LLMs in optimization. However, semantically equivalent optimizations often appear in syntactically dissimilar code, so current retrieval methods fail to identify pertinent examples, leading to suboptimal results.

To address these limitations, we propose \toolname, a framework that leverages static program analysis to identify code segments, retrieve optimization strategies, and generate optimized results. \toolname has three LLM-powered components: (1) a strategy library builder that extracts and clusters strategies from code modifications, (2) a rule generator that produces Semgrep static analysis rules to capture each strategy's applicability, and (3) an optimizer that generates optimized code using the strategy library.

On a benchmark of 151 C/C++ and 150 Python optimization tasks, \toolname shows consistent improvements across different LLMs, increasing successful optimizations by 1.38 to 28 times on C/C++ and 4.60 to 6.33 times on Python versus the baseline. On large-scale projects, \toolname improves performance metrics by 5.04\% to 218.07\% on C/C++ and 61.77\% to 479.90\% on Python, showing cross-language generalization and practical effectiveness.

\end{abstract}





\maketitle

\section{Introduction}\label{sec-intro}

Computational efficiency is a key factor in software that influences its quality~\cite{fur2011systems,garg2022deepdev}. Inefficient code segments can lead to increased system runtime, waste of computational resources, and a degraded user experience~\cite{nistor2013discovering,jovic2011catch}.
Existing research indicates that optimization opportunities are prevalent in software and may be costly and tedious for human developers to detect and optimize~\cite{jovic2011catch,dean2014perfscope}.
Therefore, automated code optimization, which involves automatically refactoring code to improve performance metrics, has attracted considerable attention from researchers in recent years.

Early research on code optimization primarily focused on rule-based approaches, which address specific types of inefficiencies~\cite{nistor2013discovering,krishna2020cadet}.
These approaches relies on predefined code-transformation rules crafted by experts. However, these rules have two fundamental limitations: (1) Due to the diversity of optimization strategies, it requires huge, if not infinite, manual efforts to craft the transformation rules for different strategies. (2) Due to the diversity of project code where an optimization strategy can be applied, it is almost impossible to write a transformation rule that capture all different cases precisely, given the expressiveness of the existing code transformation languages. 
As a result, existing rule-based optimization systems are only able to cover a narrow range of optimization problems~\cite{chen2024supersonic}.

With the advancement of deep learning technologies, particularly the emergence of large language models (LLMs), there has been a surge of research inspired by these developments and focused on LLM-based approaches~\cite{nijkamp2022codegen,li2023starcoder,guo2024deepseek,luo2023wizardcoder,xia2023keep,ye2024iter,zhang2023self,li2025editlord,fan2023large,peng2023generative,izacard2023atlas}. Compared to rule-based approaches, LLM-based approaches are able to utilize the code understanding and generation capability of LLMs, generalizing to a large variety of optimization scenarios and optimization strategies without human intervention, leading to many optimizations that are infeasible in rule-based approaches. 

\mainpoint{The Challenge of LLM-based Optimization.}
A direct approach is to give a code snippet to LLM and ask the LLM to optimize the code. However, as observed in multiple existing studies~\cite{gao2024search,shypula2023learning}, this approach is ineffective, possibly due to the many possible optimization strategies. A more effective approach is to identify the optimization strategy (e.g., replacing a linked list with an array if frequent random access is needed) to be applied in the code, and include one or a few existing optimization examples applying this strategy in the prompt for few-shot learning, such that the LLM learns the optimization strategy from the examples and applies it to the current code. 

However, applying the above few-shot learning to optimize a large software project is not easy. On the one hand, to get many optimization examples representing various optimization strategies, existing approaches~\cite{gao2024search,shypula2023learning,garg2023rapgen} usually mine optimization commits from many open source projects, forming \emph{a large space of example commits}. On the other hand, since LLM can process code only to a limited length, we have to divide the project code into multiple small chunks for LLM to process, forming \emph{a large space of code locations}. Since it is difficult to know which strategy can be applied to which code locations in advance, a trivial approach is to try all combinations of the commits and the code locations, which is infeasible in practice due to the huge number of combinations.

To address this problem, existing approaches~\cite{gao2024search,shypula2023learning,garg2023rapgen} have proposed using information retrieval techniques such as BM25~\cite{robertson1995okapi} to retrieve the most relevant optimization commit for a code location. In this way, we can enumerate the hotspot locations (which can be identified by profiling and are often small in number) and retrieve the most relevant optimization commits for these locations, effectively addressing the problem of large spaces.
However, as our evaluation will show later, since the same optimization strategy can be applied to significantly dissimilar code snippets, even the best-performing retrieval techniques in the existing studies often cannot retrieve relevant commits, leading to low performance in practice. 

\begin{figure}[t]
  \centering
  \includegraphics[width=\linewidth]{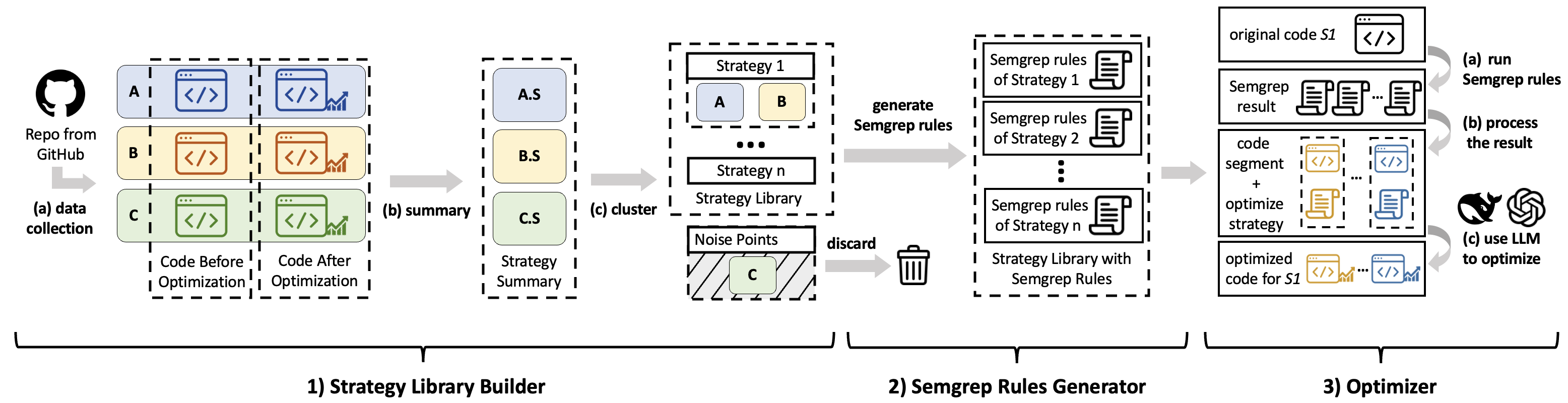}
  \caption{The overview of \toolname.}
  \label{fig:2-pipeline}
\end{figure}

\mainpoint{Our Approach.}
To address the above challenge, our main idea is that, instead of retrieving a commit suitable for the current code location, we generate a symbolic pattern-matching rule from a commit to match the potential code locations where the optimization strategy represented by the commit can be applied. More concretely, given an optimization commit, we design an LLM agent to summarize the optimization strategy used in this commit, and then generate a Semgrep~\cite{semgrep} rule to match the potential code locations that are suitable to be optimized using this strategy. Semgrep is a static analysis tool that allows the user to write customized rules to efficiently match code locations in a large project. After a potential code location is found, we ask the LLM to optimize the code using the strategy as in existing approaches. This design has three main considerations: (1) Static rules could more precisely capture the application condition of an optimization strategy, avoiding the imprecision from information retrieval techniques. (2) Utilizing the analytic power of the LLM, we can easily implement the rules for a wide range of optimization strategies, without the need of the huge human effort. (3) The static analysis rule is only used to quickly match potential code locations, and the final optimization is still decided and performed by the LLM, handling the diverse ways of applying a strategy.

With the above method, we can efficiently explore the space of code locations, but we still need to enumerate each commit in the large space of commits to generate and execute the rule extracted from this commit. To reduce this burden, we notice that many commits collected are of the same optimization strategies, and propose a clustering method to cluster the examples based on the strategies summarized by the LLM. In this way, each cluster ideally contains only commits for one strategy. Then, we produce only a few Semgrep rules from a cluster, effectively reducing the burden of enumerating the large space of examples.

However, generating Semgrep rules from optimization commits presents challenges. LLM-generated rules suffer from two problems: \emph{incompleteness}, where genuine optimization opportunities remain undetected due to rules failing to capture certain patterns, and \emph{imprecision}, where incorrect strategies are applied due to spurious matches. To address these problems, \toolname employs two complementary mechanisms: repetitive rule generation that samples multiple commits and generates multiple independent rules to expand coverage and reduce variance, and aggregation and ranking that prioritizes proposals supported by multiple independent rules to filter spurious matches. We provide theoretical analyses demonstrating that these mechanisms effectively mitigate both problems. The detailed methodology and theoretical analyses are presented in Sections~\ref{sec-approach}.

Finally, we materialize this approach into a tool called \textbf{\toolname} (for \ul{Sem}grep + \ul{Opt}imization), as shown in Figure~\ref{fig:2-pipeline}. It comprises three key components:
\begin{enumerate}
    \item A \textbf{strategy library builder} that mines optimization examples from code commits, generates their strategy descriptions by an LLM, and clusters the examples of the same strategy;
    \item A \textbf{rule generator} that generates Semgrep static analysis rules for each cluster to capture the condition of applying the optimization strategy;
    \item An \textbf{optimizer} that applies Semgrep rules on source code to detect optimizable code locations, and then uses an LLM to generate optimized code for each location.
\end{enumerate}

\mainpoint{Evaluation Result.}
To evaluate the effectiveness of \toolname, we constructed a benchmark comprising 151 C/C++ code optimization problems, covering a wide range of optimization strategies.
We then compared the performance of \toolname on this benchmark with two baseline methods using three popular LLMs: DeepSeek-V3, GPT-4.1, and Gemini-2.5-pro.
The experimental results show that \toolname achieves a significant improvement in producing correct optimization outcomes, reaching 37.5\% to 27x more successful optimizations than the baselines under different LLMs.
Furthermore, 89.86\% of the optimizations generated by \toolname preserve program semantics and yield performance improvements, demonstrating the high practical applicability of \toolname.

To evaluate whether \toolname generalizes beyond C/C++, we further constructed a benchmark comprising 150 Python code optimization tasks and conducted cross-language validation experiments using DeepSeek-V3.
The experimental results demonstrate that \toolname also significantly outperforms baseline methods on the Python benchmark, achieving improvements ranging from 58\% to 533\% in successful optimizations, confirming that the approach is not language-specific.

To evaluate the real-world performance of \toolname, we further applied \toolname to five popular open-source C/C++ projects and five popular Python projects. 
For the C/C++ projects, \toolname achieves a maximum performance improvement of 5.04\% to 218.07\% in a single performance test case, and delivers an average improvement of 1.68\% to 10.40\% across all performance test cases.
For the Python projects, \toolname achieves a maximum performance improvement of 61.77\% to 479.90\% in a single performance test case, and delivers an average improvement of 40.19\% to 143.25\% across all performance test cases.
These results demonstrate that \toolname is capable of automatically optimizing large-scale code projects across different programming languages in the real world.

\mainpoint{Our Contributions.} Our contributions are as follows.
\begin{itemize}
  \item The first LLM-based automated code optimization approach that leverages the static program analysis tool (Section~\ref{sec-approach}), which significantly improves the efficiency of identifying optimizable code snippets and selecting optimization strategy.
  \item A workflow combining LLM and clustering algorithms to summarize and merge similar optimization strategies (Section~\ref{sec-approach}), effectively reducing the size of the strategy library. 
  \item Extensive experiments on 151 C/C++ and 150 Python optimization tasks (Section~\ref{sec-empirical-eval}), demonstrating that \toolname is effective in optimizing code across different LLMs, and the improvements are consistent across programming languages, validating its cross-language generalization capability; furthermore, it can automatically optimize five large-scale C/C++ and five Python projects in the real world (Section~\ref{sec-5}), highlighting its practical value.
\end{itemize}

In the following sections, we elaborate on the design rationale and technical details of our approach. Section~\ref{sec-example} presents a systematic analysis of the problems and limitations of existing approaches in automated code optimization, discussing the limitations of existing rule-based methods, pure LLM end-to-end optimization, and retrieval-augmented methods, and introduces how \toolname addresses these problems through a combinatorial architecture. Section~\ref{sec-approach} then further details the overall framework and key components of \toolname.

\begin{figure}[t]
  \centering
  \includegraphics[width=1.0\linewidth]{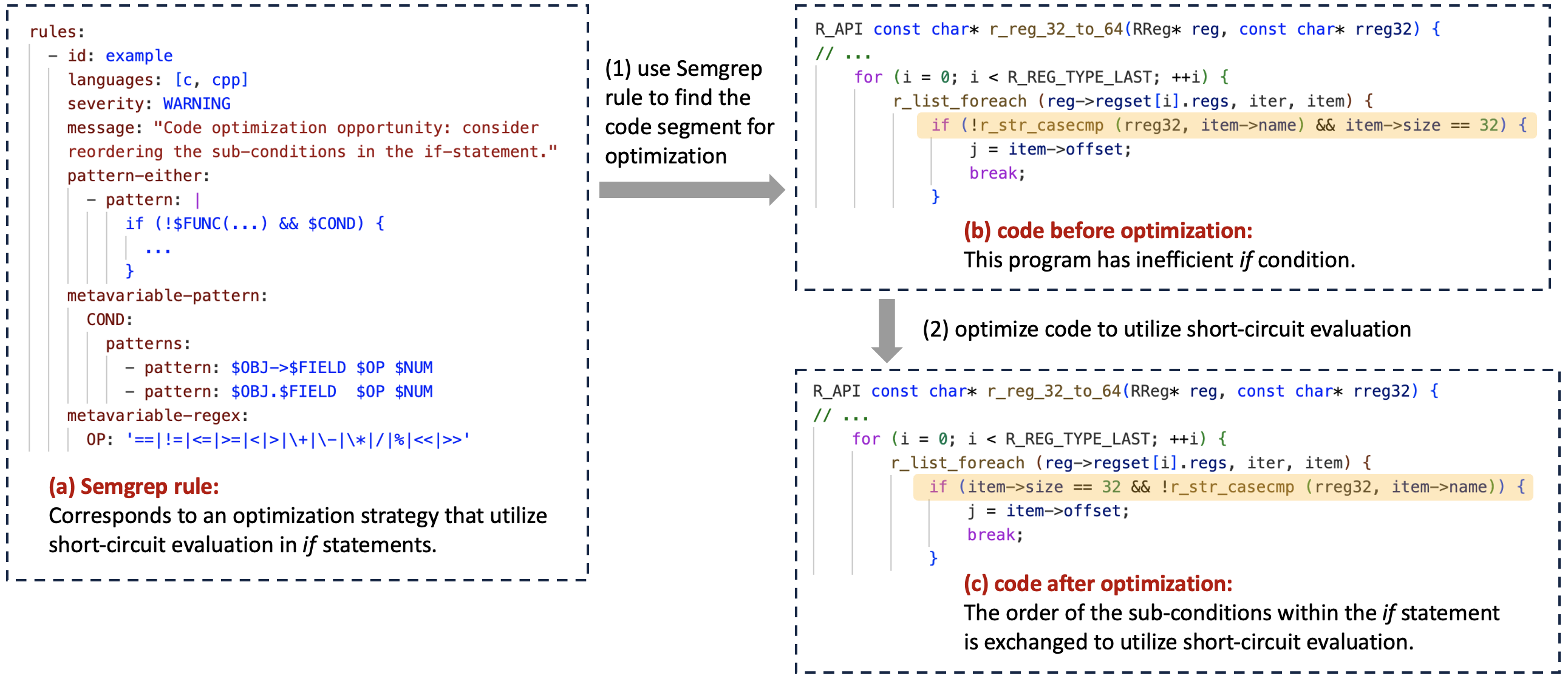}
  \vspace{-1em}
  \caption{The motivating example that leverages short-circuit evaluation to optimize the performance.}
  \label{fig:2-code_example}
\end{figure}

\section{Motivating Example}\label{sec-example}


In this section, we analyze the limitations of existing automated code optimization methods through a concrete optimization example, and thereby derive the design rationale of \toolname.

\subsection{Optimization Example Introduction}\label{sec-2-example-intro}

Figure~\ref{fig:2-code_example} shows the optimization example we use.
The code before optimization contains an \code{if} statement in a loop, which can be optimized.
The condition to be evaluated in this \code{if} statement consists of two sub-conditions connected by the \&\& operator. According to the short-circuit evaluation mechanism of Boolean expressions, if the result of the first sub-condition is false, the entire condition will be false, and thus, there is no need to evaluate the second sub-condition before proceeding to the corresponding branch.
Therefore, we can place the sub-condition that is relatively easier to evaluate at the beginning of the \code{if} statement, and the more complex sub-condition later. In this way, if the result of the earlier sub-condition is false, the program can directly proceed to the corresponding branch without evaluating the more complex part, thus improving the code execution efficiency.
In the \code{if} statement shown in Figure~\ref{fig:2-code_example}b, the evaluation of the first sub-condition is generally more complex than that of the second. Therefore, we can improve the runtime efficiency of the code by exchanging the order of the two sub-conditions.

However, existing automated code optimization methods exhibit significant limitations on this seemingly simple example. Next, we analyze in detail the shortcomings of various existing approaches in the context of this example.

\subsection{Limitations of Existing Methods}\label{sec-2-limitations}

We focus on source-level performance refactoring, rather than adjusting intermediate representations or optimization passes in compilers. The input consists of functions to be optimized along with other auxiliary information, and the output is patches that can be safely merged back into the project and bring performance improvements after compilation and testing.

Under this setting, existing automated optimization methods can be broadly categorized into three classes, each with different limitations.

\textbf{Class 1: Rule-Based Methods.} This class of methods relies on experts to manually write source-to-source transformation rules or static checking rules, then applies these rules in actual code through other toolchains to achieve code-level optimization. However, as analyzed in the introduction, these methods face two fundamental limitations: (1) Due to the high diversity of optimization strategies, crafting corresponding transformation rules for different strategies requires massive, if not infinite, manual effort. (2) Due to the high diversity of application scenarios for a single strategy in project code, the expressiveness of existing code transformation tools is limited, making it almost impossible to write rules that precisely cover all syntactic and structural variations.

Both limitations are manifested in this example. Clang-tidy~\cite{clangtidy} is a static analysis tool for C/C++ based on the Clang compiler frontend, providing code style checks and performance optimization suggestions, but it does not support optimization strategies like "swapping condition order", so it fails to optimize in this example. The CLion~\cite{clion} IDE implements built-in program transformation rules that can perform simple code optimizations. However, in testing on this example, although CLion can provide correct optimization suggestions, immediately after completing the optimization it suggests swapping the two conditions back, which precisely demonstrates the strong imprecision of its transformation rules in this scenario. That is, it completely fails to understand the semantic complexity level issue, and cannot correctly judge when to maintain the condition order and when to swap. Therefore, whenever it detects an \code{if} statement with two sub-conditions connected by the \&\& operator, it suggests the user swap the two conditions. This class of methods highly relies on experts to manually write rules, so the development cost is high and difficult to scale. Moreover, the large number of optimization strategies embodied in real commits exhibits a long-tail distribution, causing this class of methods to be unable to cover the vast majority of optimization strategies. Additionally, it is difficult for rule design to cover the various syntactic and structural variations that optimization strategies demonstrate across different projects, also leading to poor optimization effects.

\textbf{Class 2: Pure LLM End-to-End Optimization.} This class of methods directly inputs raw code along with task-related prompt information (e.g., optimization objectives) into the model, requiring the model to produce optimization results. However, since the space of potential optimization strategies is extremely large and the training data has limited coverage for each specific strategy, LLMs often can only provide generic optimization suggestions that do not improve performance (e.g., improving code readability) when facing concrete code, and struggle to identify and apply specific optimization strategies for the current scenario. This shows that, without the assistance of external tools, pure end-to-end LLM methods are unable to reliably locate and execute fine-grained optimizations for specific scenarios in such a large strategy space, and thus cannot achieve our goal of performance improvement.

\textbf{Class 3: Existing Retrieval-Augmented Methods.} This class of methods uses information retrieval techniques to retrieve knowledge potentially relevant to the current task from a knowledge base composed of historical optimization commits, then provides this knowledge along with the function to be optimized to the LLM to generate optimization patches. However, since the same optimization strategy can often be applied to code snippets with significantly different syntax and structure, text-similarity-based retrieval often fails to return truly relevant optimization examples, making it difficult for the model to learn the correct strategy applicable to the current scenario.

Taking the RAG method in this example, if we retrieve relevant commits with the code snippet in Figure~\ref{fig:2-code_example}(2), due to surface syntactic similarity, the retrieval system is more likely to return commits containing calls to {\tt r\_list\_foreach} or {\tt r\_str\_casecmp}, rather than commits adopting the "swapping condition order" strategy. This causes the retrieved examples to be irrelevant to the current optimization scenario, and the LLM fails to learn the correct optimization strategy from them, ultimately leading to optimization failure.

In summary, the three classes of methods correspond to three core challenges in automated code optimization: the manual cost and limited expressiveness of rules restrict their coverage and precision (Class 1), the overly large strategy space makes it difficult for LLMs to directly locate the correct strategy (Class 2), and similarity-based retrieval struggles to return accurate results under syntactic differences (Class 3). These limitations together show that existing methods cannot simultaneously handle strategy identification, location, and precise application, thereby motivating the design of \toolname.

\subsection{\toolname's Solution}\label{sec-2-solution}

To address these limitations, we design a combined architecture that explicitly separates strategy location from code transformation.

The optimization workflow of \toolname is shown in Figure~\ref{fig:2-pipeline}.
First, we build a library containing a large number of different code optimization strategies. Each optimization strategy is associated with corresponding Semgrep rules for static program analysis, generated automatically from commits by an LLM.
For example, there is a Semgrep rule from the strategy library in Figure~\ref{fig:2-code_example}a. It corresponds to an optimization strategy that leverages short-circuit evaluation in \code{if} statements.
Specifically, this rule matches a class of \code{if} statements containing two sub-conditions connected by an \code{\&\&} operator, with the requirement that the first sub-condition negates the result of a function call, and the second sub-condition involves accessing the value of a particular field in an object, which is then operated on with an arbitrary number.
If an \code{if} statement can be identified by this Semgrep rule, it is advisable to consider swapping the order of the two sub-conditions within the \code{if} statement to leverage short-circuit evaluation for improved code execution efficiency.

Then, whenever we have a target project to optimize, we run the Semgrep rules from the strategy library to determine at what locations the current optimization strategy can be applied. 
In Figure~\ref{fig:2-code_example}, if we run the Semgrep rule on the code before optimization, we can find that the corresponding optimization strategy can be applied, and the precise location of the \code{if} statement is identified.

\begin{figure}[t]
  \centering
  \includegraphics[width=.5\linewidth]{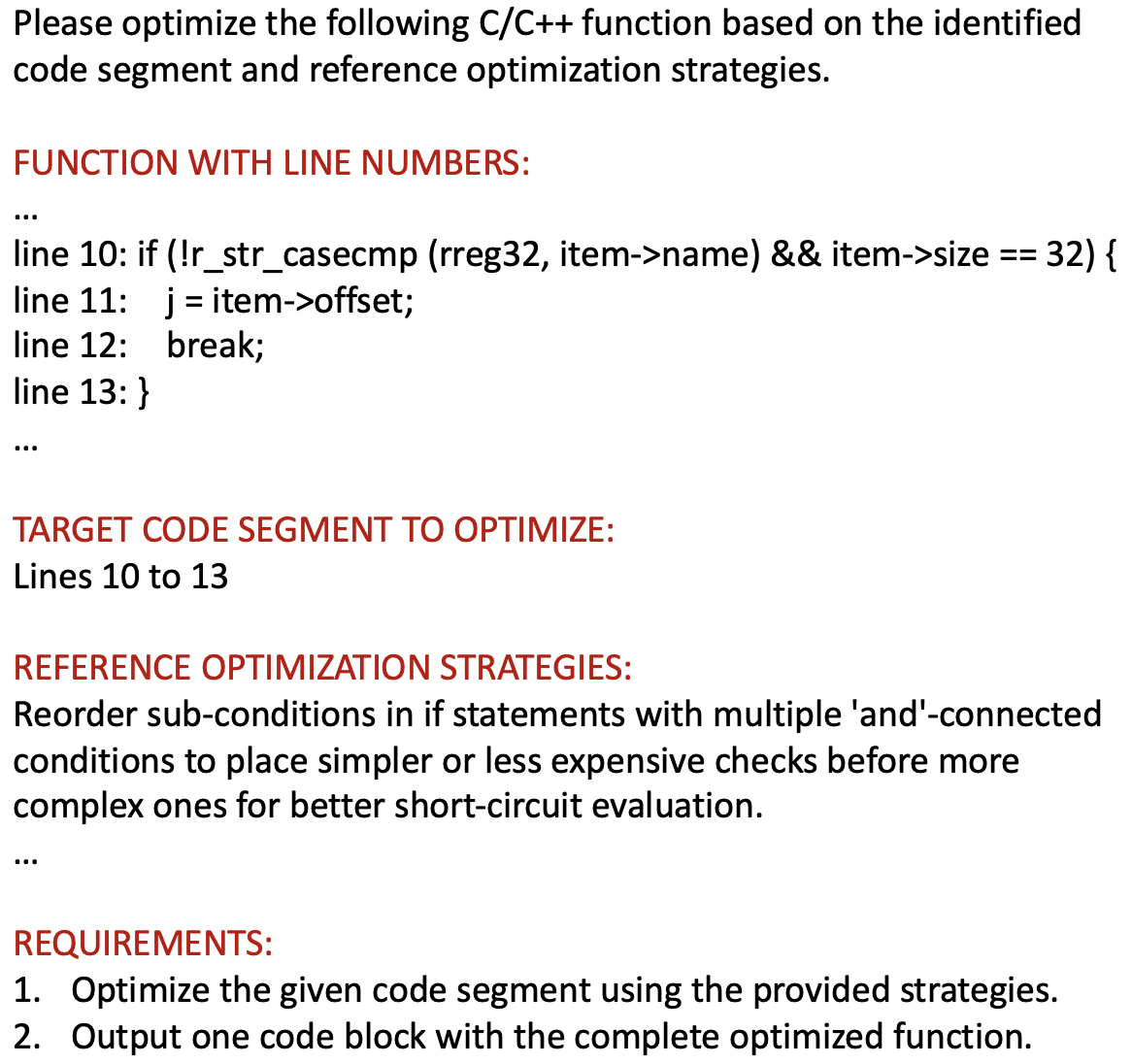}
  \vspace{-1em}
  \caption{The prompt for generating optimization results.}
  \label{fig:2-prompt}
\end{figure}

After identifying an applicable optimization strategy, we construct a corresponding optimization prompt, as shown in Figure~\ref{fig:2-prompt}.
The prompt mainly consists of four parts:
\begin{enumerate}
  \item The complete code that needs to be optimized with line number annotations
  \item The location information of the code to be optimized represented by the line number range.
  \item The optimization strategy to be used.
  \item A requirement for the model to apply the optimization strategy to the given code segment and provide the complete content of the optimized code.
\end{enumerate}

Figure~\ref{fig:2-code_example}c shows the optimization result generated by the LLM based on the prompt in Figure~\ref{fig:2-prompt}, in which the order of the sub-conditions in the \code{if} statement is swapped to utilize short-circuit evaluation and improve code execution speed. This optimization exactly matches that provided by the developer.

\subsection{\toolname's Advantages}\label{sec-2-advantages}

The design of \toolname can address the limitations of existing methods and has unique advantages.

To address the defects of rule-based or program transformer methods in terms of strategy coverage and application scenario diversity, \toolname automatically mines and clusters optimization strategies from real commits to build a large-scale strategy library, covering the diversity at the strategy level without massive manual effort. On the other hand, it leverages the LLM to generate multiple Semgrep rules for different application scenarios of each strategy, thereby covering various syntactic and structural variations of the same strategy in real code. The final concrete optimization transformation is still generated by the LLM based on the context, which can flexibly handle the diverse ways of applying the same strategy in different code, compensating for the insufficient expressiveness of fixed transformation rules.

To address the problem that pure LLM end-to-end optimization methods struggle to locate the correct strategy in a large strategy space, \toolname explicitly separates "strategy identification" from "code transformation": it first locates the code positions requiring optimization and explicitly specifies the optimization strategy through Semgrep rules, and the LLM only needs to generate code transformations under the constraints of the given strategy and location, rather than autonomously searching for which optimization strategy to use. This mechanism effectively overcomes the tendency of LLMs to generate only generic optimization proposals, enabling them to execute specific performance optimizations for concrete scenarios, and thus improves the success rate and reliability of performance optimization.

To address the defects of existing retrieval-augmented methods caused by syntactic differences leading to inaccurate retrieval results, \toolname uses Semgrep static program analysis rules to locate code positions requiring optimization, rather than relying on text-similarity-based retrieval. Semgrep rules can precisely match code syntax structures and understand semantic conditions, thereby accurately identifying code positions suitable for applying specific optimization strategies, overcoming the inaccuracy of retrieval methods under syntactic differences. Second, \toolname constructs the strategy library by automatically extracting optimization strategies from real commits, greatly expanding the optimization coverage.

In addition, existing work~\cite{yang2025knighter,liu2024write,li2025automated} has found that combining LLMs to generate correct rules for transformation is very difficult, and it is hard to correct rules once they are wrong.
However, \toolname adopts a two-part separated architecture.
The first part is responsible for using Semgrep rules to construct a candidate set with high recall across the strategy and location spaces, that is, increasing the probability of containing the correct optimization scheme.
The second part contains multi-layer filtering mechanisms and uses LLM to generate optimization results. The multi-layer filtering mechanism selects the most credible $N$ optimization schemes from scan results. If the LLM finds that a piece of code is not suitable for optimization using the given scheme, it will not make any modifications. Both the filtering mechanism and LLM can ensure that precision is improved while not excessively reducing recall, that is, reducing incorrect optimizations.
Therefore, \toolname can appropriately relax the requirements for Semgrep rule precision, reducing the difficulty of generating Semgrep rules, which solves the problem that directly generating rules in existing work is too difficult.

Note that the rules in \toolname are not necessarily precise (e.g., in this example, a function call may not necessarily lead to a heavier evaluation than the second condition) or complete (e.g., the example optimization strategy can be applied even if there is no negation), since they are generated by an LLM. To address these issues, we further employ two methods: 1) to address \emph{incompleteness}, we generate multiple rules per strategy, where the rules together forming a more complete set with higher recall; 2) to address \emph{imprecision}, we aggregate and sort the locations by the number of rules matching them, where the location matched by more rules are more likely to be true optimization opportunities, and the method mentioned above also addresses this problem. The detailed methodology is presented in Section~\ref{sec-approach}, and the theoretical analyses demonstrating how these mechanisms effectively mitigate both problems are provided in Section~\ref{sec-3-4}. Furthermore, the precision issue is also addressed by the later optimization generation step, as the LLM will fail to generate an optimized code snippet if the code is unsuitable to be optimized using this strategy. 

The subsequent experimental sections (Section~\ref{sec-empirical-eval} and Section~\ref{sec-5}) conduct systematic evaluations on both C/C++ and Python versions of \toolname. Through comparison with several baseline methods, we verify that the tool can effectively improve optimization success rates across multiple benchmarks and real-world projects, while demonstrating good cross-language applicability.

The example in this section is intentionally simple for clarity of \toolname's overall workflow.
In practice, \toolname is capable of handling more complex optimization scenarios with domain-specific strategies. A more complex optimization example is presented in Section~\ref{sec-5-example}, and all related work mentioned above fails to produce correct optimization results on this example, further validating the effectiveness of \toolname.

\section{Approach}\label{sec-approach}

In this section, we present the details of three components in \toolname: a library builder in Section~\ref{sec-3-1}, a rule generator in Section~\ref{sec-3-2}, and an optimizer in Section~\ref{sec-3-3}. We then provide theoretical analyses of key mechanisms in \toolname in Section~\ref{sec-3-4}.
The overall pipeline of our approach is illustrated in Figure~\ref{fig:2-pipeline}.

\subsection{Building the Strategy Library}\label{sec-3-1}

The first step of our methodology is to construct a code optimization strategy library, which contains textual descriptions and representative code examples for each optimization strategy. Subsequently, for each optimization strategy, we generate corresponding Semgrep rules (Section~\ref{sec-3-2}), thereby establishing a comprehensive strategy library.
The static program analysis rules in this strategy library can be used to identify optimization opportunities, which then guide the LLM in applying the corresponding strategies and producing optimized code. The following describes how relevant data is collected to construct the strategy library, as well as how clustering is employed to retain effective information and reduce the size of the library.

\subsubsection{Data Collection}\label{sec-3-1-1}

The first step of our approach is to collect a large set of optimization commits.
The data collection process of \toolname is language-agnostic and can be instantiated for any language supported by Semgrep.
We have currently implemented two versions: the first version processes C/C++ code, and the second version processes Python code.
Below we describe the instantiation data for both versions.

\textbf{C/C++ Version.}
We collected 30,342 codebases from GitHub with at least 100 stars, whose primary language is C or C++, and which have had commit activity within the past five years.
We crawled the commit histories of the main branches of these codebases, ensuring that for each commit, we collected the commit message, the complete code before the modification, the complete code after the modification, and the diff file that records the specific changes introduced by the commit. In total, we obtained 108,039,931 commits.

In this work, we focus solely on code optimizations within individual functions. Therefore, we first filter for commits that modify only a single function in C or C++.
Next, we employ a two-step filtering process to identify commits related to code optimization. First, we use a set of optimization-related keywords (listed in Appendix~\ref{sec-9-1}) to match commit messages, preliminarily filtering out commits that may involve code optimization. Then, for these candidates, we incorporate additional information, including the repository, commit message, and code changes, and utilize an LLM (with the prompt template provided in Appendix~\ref{sec-9-2-1}) to further determine whether the commit primarily implements code optimization.
Through this process, we ultimately identify 35,668 commits that primarily implement code optimizations, which are used to construct the C/C++ strategy library.

\textbf{Python Version.}
We follow the same data collection process as the C/C++ version.
We collected 9,316 codebases from GitHub with at least 100 stars, whose primary language is Python, and which have had commit activity within the past five years, and crawled the commit histories of the main branches of these codebases.
In total, we obtained 14,821,894 commits.
Through the same filtering process, we ultimately identify 10,084 commits that primarily implement code optimizations, which are used to construct the Python strategy library.

\subsubsection{Summarization of Optimization Strategies}\label{sec-3-1-2}

This part is mainly divided into two modules.

First, we use an LLM to generate summaries of the optimization strategy used in each commit. Concretely, for each code optimization-related commit, we collect detailed information, including the name of the codebase, the commit message, the name of the modified function, and the complete diff file.
Then, we include all this information in the prompt and use LLMs to generate a one-sentence summary of the corresponding optimization strategy. To reduce variability in the generation process and improve the reliability of the results, we independently generate $m$ summaries for each commit. In our implementation, we set $m$ as 3.

Second, we select the most representative summary among the generated summaries for each commit. Specifically, we first use a pre-trained model to encode each summary into a high-dimensional semantic vector. Next, we calculate the pairwise cosine similarities to construct a similarity matrix. Finally, we compute the average similarity of each summary to the others and select the one with the highest average as the final optimization strategy summary for that commit.

The complete prompt template used in this phase is provided in Appendix~\ref{sec-9-2-2}.

\subsubsection{Strategy Integration and Selection}\label{sec-3-1-3}

First, as in previous steps, we utilize a pre-trained model to convert the strategy summary corresponding to each commit into a high-dimensional semantic vector.

Then, we apply an unsupervised clustering algorithm, DBSCAN~\cite{ester1996density}, based on cosine similarity to all semantic vectors. By setting the hyperparameter $eps$ as $0.89$ for C/C++ code and $0.85$ for Python code (both determined via pilot study in Section~\ref{sec-4-pilot-e1}), we control the minimum similarity within each cluster, ensuring that the strategy summaries grouped together exhibit a high degree of semantic similarity in the vector space.
Additionally, we set the minimum cluster size as 5 (determined via pilot study in Section~\ref{sec-4-pilot-e1}), as clusters with insufficient size may represent strategies with limited practical value. Removing them could improve the quality of the library and the efficiency of the whole system. Finally, we consider each cluster to represent one optimization strategy.

The detailed clustering statistics are shown in Table~\ref{tab-clustering-statistics}.

\begin{table}[htbp]
\centering
\caption{Strategy Library Clustering Statistics}
\label{tab-clustering-statistics}
\begin{adjustbox}{max width=\linewidth}
\begin{tabular}{
    >{\centering\arraybackslash}m{3.5cm}
    >{\centering\arraybackslash}m{3.5cm}
    >{\centering\arraybackslash}m{3.5cm}
}
\toprule
\textbf{Metric} & \textbf{C/C++ Version} & \textbf{Python Version} \\
\midrule
Total commits & 27,463 & 10,073 \\
Number of clusters & 155 & 55 \\
Clustering rate & 7.54\% & 10.87\% \\
Max cluster size & 146 & 159 \\
Min cluster size & 5 & 5 \\
Avg cluster size & 13.38 & 19.91 \\
Median cluster size & 6 & 7 \\
Std deviation & 22.09 & 32.40 \\
\bottomrule
\end{tabular}
\end{adjustbox}
\end{table}

\subsection{Automatic Generation of Analysis Rules}\label{sec-3-2}

The rule generation component of \toolname is built on Semgrep. Semgrep supports multiple programming languages (including C/C++, Python, Java, etc.), and its rule syntax is language-independent.
Therefore, the rule generation process described in Sections~\ref{sec-3-2-1} and~\ref{sec-3-2-2} can be instantiated to any language supported by Semgrep by switching the target language and rule templates in the prompts.

After constructing the basic optimization strategy codebase using the method described in Section~\ref{sec-3-1}, we generate corresponding Semgrep rules for each category of optimization strategy to identify potential optimization opportunities. In particular, multi-commit sampling in Section~\ref{sec-3-2-1} and repeated independent rule generation per commit in Section~\ref{sec-3-2-2} help address \emph{incompleteness} and \emph{imprecision}.

\subsubsection{Select Commits for Each Cluster}\label{sec-3-2-1}

As shown in Section~\ref{sec-example}, a single commit may not fully represent all contexts in which an optimization strategy can be applied, and the rule generated from it may also be affected by commit-specific noise. 
To address these issues, \toolname randomly samples up to $n$ commits from each strategy cluster (we set $n=10$ in practice, determined via pilot study in Section~\ref{sec-4-pilot-e2}) and generates one Semgrep rule from each sampled commit. 
This design provides two complementary benefits.

First, different commits often apply the same optimization strategy in distinct syntactic or semantic contexts. 
By sampling multiple commits, \toolname expands the observable space of code patterns where the strategy is applicable, thereby increasing coverage and alleviating potential \emph{incompleteness} that may arise when relying on a single commit.

Second, sampling multiple commits for each strategy helps mitigate the instability and noise that may be introduced by individual commits, which alleviates \emph{imprecision}. 

For a detailed theoretical analysis of this variance reduction effect, please refer to Section~\ref{sec-3-4-1}.

\subsubsection{Generate Semgrep Rules for Each Commit}\label{sec-3-2-2}

For each commit, we have developed a dedicated agent that automatically generates Semgrep rules by iteratively invoking an LLM and incorporating its feedback. The main process is as follows:

\begin{itemize}
  \item \textbf{Understand the optimization strategy:} We provide the diff file of the commit to the LLM, and request it to analyze and explain the optimization strategy used in this commit. At this stage, the focus is solely on understanding; no Semgrep rules are generated.
  \item \textbf{Generate an initial rule:} Based on the previous analysis, the LLM produces a candidate Semgrep rule. This rule should be designed to detect similar optimization opportunities and must strictly adhere to Semgrep's syntax requirements.
  \item \textbf{Validate and iteratively refine the rule:} The generated Semgrep rule is executed on the pre-commit version of the code, where the commit is used for rule generation. If an error occurs, the error information and context, together with the current rule, are automatically fed back to the LLM for targeted revision. This process repeats until the rule executes successfully or a predefined maximum number of iterations is reached.
  Specifically, we configure Semgrep to output results in JSON format and mainly handle two types of errors. The first type is a complete execution failure, where no valid JSON output can be obtained; in this case, the LLM is directly informed that the execution did not produce a valid result and a revision is required. The second type refers to cases where a JSON file is generated, but it contains a non-empty \texttt{"error"} or \texttt{"fatal"} field. For such cases, we extract the specific error information and provide it to the LLM for targeted modification. Common issues such as syntax errors and malformed rules are typically categorized as errors.
\end{itemize}

The complete prompt templates used in this phase are provided in Appendix~\ref{sec-9-2-3}.

It should be noted that our iterative refinement process only requires that the generated Semgrep rules can execute without errors, without requiring the rules to be functionally perfect. This is because, as discussed in Section~\ref{sec-example}, in the \toolname framework design, Semgrep rules are solely responsible for locating potential optimization opportunities, and the final optimization decisions are made jointly by subsequent multi-layer mechanisms (including aggregation and ranking, Top-N selection, and LLM optimization). Even if Semgrep rules have certain false positives, subsequent stages can perform multi-layer filtering and judgment to correct possible errors and improve overall accuracy. Therefore, we have relatively relaxed requirements for rule precision, focusing more on improving recall.

Additionally, based on practical experiments and insights from existing work~\cite{gao2023makes,yang2025knighter,liu2024write}, we observed that multi-round iterative refinement often performs worse than regeneration from scratch. This is because during the iterative refinement process, the error context from the previous round accumulates, making it difficult for the LLM to break free from its original mindset. Related works also report similar findings, i.e., iterative refinement struggles to fix fundamental errors. Therefore, \toolname generates Semgrep rules independently for each commit $t$ times (we set $t=5$ in practice, determined via pilot study in Section~\ref{sec-4-pilot-e2}), with each generation executed from scratch and allowed up to seven refinement iterations.

This design helps mitigate errors or biases that may arise from a single generation attempt, which reduces \emph{imprecision}. 
Moreover, each independently generated Semgrep rule may capture different aspects of the optimization strategy or be applicable to different code contexts, thereby alleviating \emph{incompleteness}. 
For a detailed theoretical analysis of this probabilistic coverage effect, please refer to Section~\ref{sec-3-4-1}.

\subsubsection{Statistical Result}\label{sec-3-2-3}

After the clustering and rule generation process described above, the \toolname strategy library comprises a total of $155$ C/C++ optimization strategy clusters and $55$ Python optimization strategy clusters. Detailed statistics are shown in Table~\ref{tab-rule-generation-statistics}.

\begin{table}[htbp]
\centering
\caption{Rule Generation Statistics}
\label{tab-rule-generation-statistics}
\begin{adjustbox}{max width=\linewidth}
\begin{tabular}{
    >{\centering\arraybackslash}m{4.5cm}
    >{\centering\arraybackslash}m{2.8cm}
    >{\centering\arraybackslash}m{2.8cm}
}
\toprule
\textbf{Metric} & \textbf{C/C++ Version} & \textbf{Python Version} \\
\midrule
Total Clusters & 155 & 55 \\
Processable Commits & 1,107 & 420 \\
Theoretical Rules to Generate & 5,535 & 2,100 \\
Successfully Generated Rules & 5,057 & 2,097 \\
Rule Success Rate & 92.28\% & 99.86\% \\
\bottomrule
\end{tabular}
\end{adjustbox}
\end{table}

This scale demonstrates the library's broad coverage and the diversity of code optimization strategies incorporated by \toolname.

\subsection{Strategy Library-Based Code Optimization}\label{sec-3-3}

The code optimization workflow of \toolname is independent of the specific programming language. As long as Semgrep can provide matching locations and the LLM can generate code in the target language, the optimizer can function properly.
This design enables the same optimization workflow to be directly applied to different languages such as C/C++ and Python without modifying the core logic.

After generating Semgrep rules for each strategy in the strategy library using the method described in Section~\ref{sec-3-2}, the library can then be utilized to optimize code. This section presents the code optimization workflow.

First, after obtaining code that needs to be optimized, we apply all Semgrep rules associated with the strategies in the library to scan the code and record the raw matches.
Each match captures two key pieces of information: the precise location of the matched code fragment and the optimization strategy associated with the firing rule.
Since Semgrep is a static program analysis tool, it can analyze incomplete code (e.g., individual files or functions) and operates with high speed, typically completing all scans on files with several hundred or even over a thousand lines of code in about 1 second.

Next, we aggregate and rank the scan results as optimization proposals. Since each type of strategy may correspond to multiple Semgrep rules, the same location may be matched by multiple rules. As mentioned in Section~\ref{sec-example}, since the rules may be imprecise, we can use the number of rules matching a location as an indicator to show how likely the location is a true optimization opportunity.
Each proposal specifies which strategy should be applied at which location—that is, a pair consisting of a concrete code location and an optimization strategy. We merge identical proposals across all rules: if multiple Semgrep rules originating from the same strategy report the same code fragment, we collapse them into one proposal and maintain an occurrence count for that proposal. Then, for each function, we sort proposals by their counts in descending order and retain only the top $n$ (we set $n=25$ in our implementation, determined via pilot study in Section~\ref{sec-4-pilot-e9}).

Finally, for each selected proposal, we compile all relevant context and invoke the LLM to generate the optimized result. As illustrated in Figure~\ref{fig:2-prompt}, the prompt includes: the location of the code fragment to be optimized, the optimization strategy to be applied, and the complete code requiring optimization. For further details, please refer to Section~\ref{sec-example}. The complete prompt template is provided in Appendix~\ref{sec-9-2-4}.

In the second step, ranking optimization proposals by the number of matching rules helps filter out spurious matches from rule generation errors or imprecise patterns, while proposals supported by multiple independent Semgrep rules are more likely to represent genuine optimization opportunities, thereby mitigating \emph{imprecision}. For a detailed theoretical analysis of this aggregation mechanism, please refer to Section~\ref{sec-3-4-2}.

\subsection{Property of \toolname}\label{sec-3-4}

This section provides theoretical analyses of key mechanisms in \toolname to demonstrate their effectiveness in addressing \emph{incompleteness} and \emph{imprecision}.

Section~\ref{sec-3-4-1} analyzes how the repetitive mechanisms employed in Sections~\ref{sec-3-2-1} and~\ref{sec-3-2-2} mitigate \emph{incompleteness} and \emph{imprecision} through variance reduction.
Section~\ref{sec-3-4-2} analyzes how the aggregation and ranking mechanism employed in Section~\ref{sec-3-3} mitigates \emph{imprecision}.

\subsubsection{Theoretical Analysis of Repetitive Mechanisms}\label{sec-3-4-1}

As described in Sections~\ref{sec-3-2-1} and~\ref{sec-3-2-2}, \toolname employs two repetitive mechanisms to generate Semgrep rules: sampling multiple commits from each strategy cluster and generating multiple independent rules for each commit.
This section analyzes how these mechanisms jointly alleviate \emph{incompleteness} and \emph{imprecision}.

\paragraph{Mitigating Incompleteness via Coverage Expansion.}

Let $p_i$ denote the probability that the $i$-th Semgrep rule correctly captures a true optimization opportunity.
Assuming independence across different rule generation processes, the coverage probability for $k$ rules can be expressed as:
\[
P_{\text{coverage}}(k) = 1 - \prod_{i=1}^{k} (1 - p_i),
\]
which represents the probability that at least one of the $k$ rules successfully detects the optimization opportunity.
When $p_i > 0$ for all $i$, $P_{\text{coverage}}(k)$ increases monotonically with $k$.
Each rule generation may capture different syntactic or semantic manifestations of the same strategy or be applicable to different code contexts, and generating multiple rules expands the collective coverage.

Therefore, the repetitive mechanisms effectively reduce \emph{incompleteness} by generating multiple rules for each strategy, which increases the coverage probability.

\paragraph{Mitigating Imprecision via Variance Reduction.}

Next, we analyze the variance reduction effect.
Let $x$ denote a code location where an optimization strategy might be applicable, and let $k$ denote the number of Semgrep rules generated for that strategy.
For the $i$-th rule, let $\hat f_i(x)\in\{0,1\}$ denote the output at location $x$, where $i \in \{1, 2, \ldots, k\}$.
We define the expected value as
\[
\mu(x)\;=\;\mathbb{E}[\hat f_i(x)],
\]
which represents the probability that a randomly selected rule from the generation process would match at location $x$.
The per-location variance component is defined as
\[
\mathrm{Var}_x \;=\; \mathbb{E}\big[(\hat f_i(x) - \mu(x))^2\big].
\]

When analyzing the aggregation effect of multiple rules, we consider the average match frequency \(\bar f(x)=\frac{1}{k}\sum_{i=1}^k\hat f_i(x)\) and approximate its variance using ensemble analysis~\cite{breiman1996bagging}.
Under standard ensemble-analysis assumptions (approximately equal marginal variance across rules and approximately homogeneous pairwise correlations), the variance admits the following approximation:
\[
\mathrm{Var}\big[\bar f(x)\big] \approx \sigma_v^2(x)\!\left(\rho(x) + \frac{1-\rho(x)}{k}\right).
\]
Here, $\sigma_v^2(x)=\mathrm{Var}_x$ is the single-rule variance at location $x$, and $\rho(x)\in[-1,1]$ is the average pairwise correlation coefficient between rule outputs at that location.

Since each rule generation process is executed independently from scratch, the resulting $k$ rules can be treated as approximately independent and identically distributed with respect to the generation randomness.
Therefore, the approximation formula above can be applied to quantify the variance reduction effect.

When the correlation between rules is low, i.e., $\rho(x)$ is close to 0, the formula simplifies to $\mathrm{Var}\big[\bar f(x)\big] \approx \sigma_v^2(x) / k$, meaning that by increasing the number of rules $k$, \toolname can reduce the aggregated variance by a factor of $1/k$.
Additionally, by generating multiple independent rules for the same commit and sampling across different commits, \toolname can effectively reduce the correlation between rules, further decreasing the variance.

In summary, the repetitive mechanisms effectively mitigate \emph{imprecision} by generating multiple independent rules for each strategy, which reduces the aggregated variance with $1/k$ scaling and makes the rule set for each strategy more stable.

\subsubsection{Theoretical Analysis of Aggregation and Ranking Mechanism}\label{sec-3-4-2}

As described in Section~\ref{sec-3-3}, \toolname aggregates and ranks optimization proposals by the number of matching Semgrep rules, then selects the top-ranked proposals for code generation.
This section analyzes how this ranking mechanism mitigates \emph{imprecision} in the final optimization process.

To formally analyze the reliability improvement achieved by aggregated rule ranking, we model the rule-matching process as a probabilistic ensemble of independent binary detectors.
For a given code location and optimization strategy, let there be $k$ Semgrep rules associated with the strategy.
Let the random variable $M$ denote the total number of rules that report a match for that location.

We define two key conditional probabilities that characterize the behavior of a single rule:
\[
p = P(\text{rule matches} \mid \text{true optimization opportunity}), 
\quad
q = P(\text{rule matches} \mid \text{non-optimization opportunity}),
\]
where $p$ is the true detection probability and $q$ is the false detection probability.
A reliable rule satisfies $p>q$.

Under the independence assumption among rules, the number of matches $M$ follows a binomial distribution: for genuine optimization opportunities, $M \sim \mathrm{Binomial}(k,p)$; for false positives, $M \sim \mathrm{Binomial}(k,q)$.
When \toolname ranks code locations by the number of matching rules, applying a threshold $r$ corresponds to accepting locations with $M \ge r$.
The true positive rate (TPR) and false positive rate (FPR) are thus:
\[
\mathrm{TPR}(r) = \sum_{i=r}^{k} \binom{k}{i} p^{i} (1-p)^{k-i}, \qquad
\mathrm{FPR}(r) = \sum_{i=r}^{k} \binom{k}{i} q^{i} (1-q)^{k-i}.
\]

Since $p>q$, $\mathrm{FPR}(r)$ decreases at a faster rate than $\mathrm{TPR}(r)$ as $r$ increases.
This asymmetric decay reflects a fundamental property of ensemble classifiers: aggregating multiple independent detectors amplifies the statistical difference between true and false matches.

We further consider the ratio $R(r) = \mathrm{FPR}(r) / \mathrm{TPR}(r)$, which measures the relative proportion of false positives among accepted proposals.
Since $\mathrm{FPR}(r)$ decays faster than $\mathrm{TPR}(r)$, $R(r)$ decreases monotonically with $r$, demonstrating that higher thresholds yield higher reliability.

This probabilistic model aligns with the standard framework of threshold-based ensemble classification~\cite{breiman1996bagging}, where the consistency of multiple independent predictions serves as evidence of correctness.
Accordingly, the ranking mechanism in \toolname—prioritizing locations supported by more matching rules—naturally implements a confidence-weighted ensemble filter: higher match counts correspond to higher confidence, which suppresses spurious matches from noisy or imperfect rules.

Optimization proposals ranked higher in the aggregated list possess higher statistical confidence, while those ranked lower are less reliable.
To balance reliability and coverage, \toolname selects only the top-ranked proposals for optimization, ensuring that most true opportunities are retained while effectively filtering out unreliable cases.

In summary, the aggregation and ranking mechanism employed in Section~\ref{sec-3-3} effectively mitigates \emph{imprecision} by prioritizing optimization proposals supported by multiple independent Semgrep rules, which reduces the false positive rate more rapidly than the true positive rate and thereby filters out spurious matches from rule generation errors or imprecise patterns.


\section{Empirical Evaluation}\label{sec-empirical-eval}

We evaluate \toolname with the following research questions:

\begin{enumerate}[left=1.5em]
    \item[\textbf{RQ1:}] Does \toolname produce more successful optimizations than baselines on both C/C++ and Python benchmarks?
    \item[\textbf{RQ2:}] How important is each component of \toolname?
    \item[\textbf{RQ3:}] Do optimization strategies in \toolname and baselines generalize across codebases?
    \item[\textbf{RQ4:}] What types of optimization strategies does the \toolname strategy library cover, and what are the success rates for each category?
    \item[\textbf{RQ5:}] What are the main failure modes of \toolname and what proportions do they occupy?
    \item[\textbf{RQ6:}] How effective are the Semgrep code location and ranking mechanisms in \toolname?
\end{enumerate}

\subsection{Experimental Setup}\label{sec-4-setup}

\subsubsection{Benchmark}

In this work, we constructed benchmarks for both C/C++ and Python to evaluate the cross-language effectiveness of code optimization tools. Both benchmarks follow the same construction process: collecting commits from the 100 most-starred codebases for each language, filtering commits that modify only a single function, identifying optimization commits through keyword filtering and LLM verification, and randomly sampling after deduplication.

\vspace{0.5em}
\noindent\textbf{C/C++ Benchmark.} We collected 2,953,660 commits from the main branches of the 100 most-starred C/C++ codebases on GitHub.
We first filtered commits to retain only those modifying a single C/C++ function. Using optimization-related keywords to match commit messages, we then employed an LLM to analyze the corresponding diff files for verification.
After deduplication based on commit messages and code changes, we obtained 2,529 commits that primarily implement code optimizations, from which we randomly sampled 151 commits to construct our benchmark dataset.

\vspace{0.5em}
\noindent\textbf{Python Benchmark.} We collected 1,083,554 commits from the main branches of the 100 most-starred Python codebases on GitHub.
Following the same filtering process as C/C++, we obtained 1,114 commits that primarily implement code optimizations, from which we randomly sampled 150 commits to construct our Python benchmark dataset.
This Python benchmark is primarily used to validate the generalization capability of \toolname across different programming languages.

Since the benchmarks are constructed from historical commits on GitHub, they overlap with the data used by baselines and \toolname. This includes the knowledge base employed by RAPGen and RAG, as well as the strategy library of \toolname.
To address the data leakage problem, during the evaluation, we exclude information exactly matching the commit or the code of the current task from the data used in baselines and \toolname. Only the remaining data is available for optimization, thereby mitigating the problem of data leakage.

Each entry in the benchmark contains the codebase name, the commit hash, and the complete function before and after the commit. The function before commit is the input for code optimization tools, and the function after commit is used to calculate the ground truth patch by the developer, so that we can assess the correctness of the optimization results.

Note that we constructed a new benchmark instead of reusing existing ones because no existing benchmarks suit our needs as far as we are aware. For example,
the benchmark used by RAPGen~\cite{garg2023rapgen} consists of optimization tasks for C\# code, whereas our tool is implemented to optimize C/C++ and Python code.
The PIE benchmark~\cite{shypula2023learning} derived from CodeNET~\cite{puri2021codenet} is typically small in scale, whereas we aim to evaluate optimization capabilities on large-scale codebases in real-world scenarios.

\subsubsection{Baselines}

We compared \toolname with three baseline approaches on both C/C++ and Python benchmarks. For the C/C++ benchmark, we additionally include the traditional static analysis tool Clang-tidy as a baseline. Since Clang-tidy does not support Python, this baseline is only evaluated on the C/C++ benchmark.

\mainpoint{Direct Prompting:} This baseline uses a static prompt, requiring the LLM to directly provide the optimized results.

\mainpoint{Retrieval Augmented Generation (RAG):}
This baseline implements the standard RAG process.
The implementation follows previous work~\cite{gao2023makes}. We first gather real-world code optimization-related commits to construct a comprehensive knowledge base. The collection process follows the same approach as in Section~\ref{sec-3-1}. For each piece of code to be optimized, we follow \citet{gao2023makes} to use the BM25 algorithm~\cite{robertson1995okapi} to retrieve the four most similar code examples from the knowledge base, sort them in ascending order of similarity, and then include them in the prompt along with the current code to be optimized. The LLM is then instructed to generate an optimized version of the code.

\mainpoint{RAPGen:} We compared against RAPGen~\cite{garg2023rapgen}, the state-of-the-art approach for large-scale code optimization as we are aware.
Since RAPGen's source code is not publicly available, we re-implemented the tool according to the descriptions in its paper and adapted it for code optimization tasks on C/C++ and Python code.
Since the RAPGen approach does not identify the location to be optimized, it is necessary to specify the line number of the code to be optimized in the input. The RAPGen then uses the corresponding code line to match optimization strategies within its knowledge base. If multiple strategies are found, the tool calculates similarity scores to identify the most suitable optimization strategy, which is then applied to the code.
In our implementation of the RAPGen tool, it enumerates each line of the code and attempts to match optimization strategies from the knowledge base for each line. After collecting all potential strategies, we then use RAPGen's similarity calculation method to identify the final strategy to be applied.

Additionally, we also implemented an enhanced version of the RAPGen tool (RAPGen+), which additionally provides the location information of the modified code segments from the original commit as input. Subsequently, only the code within these segments is used to match optimization strategies in the knowledge base, after which the similarity is calculated to determine the final strategy to be applied. This setup represents an ideal upper bound of the performance of RAPGen if the optimization location is precisely known in advance (which is not true for all other baselines).

\mainpoint{Clang-tidy:} Clang-tidy~\cite{clangtidy} is a Clang-based C++ static analysis tool primarily used for code style checking and simple optimizations. It detects inefficient patterns in code through predefined rules and attempts to provide automated fix suggestions.

\subsubsection{Language Coverage in Experiments}

In this experiment, RQ1 is evaluated on both C/C++ version and Python version to verify the cross-language generalization capability of \toolname. However, RQ2 to RQ6 are conducted only on the C/C++ version.

\subsubsection{Evaluated LLMs}
\label{sec-4-llms}
Our evaluation of \toolname and the baselines was performed on three LLMs, i.e., DeepSeek-V3 (version 0324)~\cite{liu2024deepseek}, GPT-4.1 (version 2025-04-14)~\cite{openai2025gpt41}, and Gemini-2.5-Pro~\cite{deepmind2025gemini25pro}, via their official APIs. Our choice of LLMs covers both open-source and proprietary LLMs with decent coding and reasoning capabilities.
Due to cost constraints, we evaluated Gemini-2.5-Pro only for direct prompting, RAG, and \toolname on a randomly selected subset of 40 code optimization problems. For the ablation studies in RQ2, we exclusively used DeepSeek-V3.

For Python experiments, we used DeepSeek-V3 only, as the goal is to validate cross-language effectiveness rather than cross-model performance. This approach also allows direct comparison with C/C++ results under the same LLM.

\subsubsection{Metrics}

To evaluate the correctness of the optimized code, we employ two metrics, following existing work~\cite{garg2023rapgen}.

\begin{itemize}
    \item \textbf{Exact Match (EM):} We normalize both the ground-truth code by the developer and the optimization results generated by each tool by removing comments, indentation, and whitespace characters. We then compare the normalized code strings to check whether they are identical.
    \item \textbf{Semantic Equivalence (SemEqv):} We retrieve the automatically generated patch and the developer's ground-truth patch. The authors then manually compare the two patches to determine whether they are semantically equivalent.
\end{itemize}

For each optimization task, if at least one solution generated by a given method meets the evaluation criteria, we consider the method to have successfully solved the task.

Note that we do not use CodeBLEU~\cite{ren2020codebleu} metrics in our evaluation. We employ manual evaluation in the Semantic Equivalence metric, which is more direct and rigorous than automated proxy metrics such as CodeBLEU. While RAPGen~\cite{garg2023rapgen} employs CodeBLEU to approximate manual evaluation and reduce manual labor, it also relies on manual evaluation as the superior metric where experts assess optimization suggestions. Since we manually evaluate all results for the Semantic Equivalence metric, CodeBLEU as a proxy for manual evaluation is unnecessary.

\subsubsection{Implementation Details}

For all LLMs, we follow prior work~\cite{gao2023makes,cheng2022binding,nashid2023retrieval} and set the temperature to 0. Each optimization task is repeated three times, and we report the average number for all metrics to mitigate randomness. 
To build the strategy library (Section~\ref{sec-3-1}), we use DeepSeek-V3~\cite{liu2024deepseek} to summarize the optimization strategy, and all-MiniLM-L6-v2~\cite{wang2020minilm,allminilm2021} to encode the summary into a high-dimensional semantic vector.

\subsection{Pilot Study: Parameter Configuration}\label{sec-4-pilot}

Before conducting the main experiments, we first perform a pilot study to determine the optimal values for key hyperparameters in our approach, providing empirical evidence for the clustering parameters in Section~\ref{sec-3-1-3}, the commit sampling number in Section~\ref{sec-3-2-1}, and the candidate retention number in Section~\ref{sec-3-3}.

\subsubsection{Overview}\label{sec-4-pilot-overview}

This pilot study uses a dataset completely independent from the main experiments to ensure the objectivity and generalizability of parameter selection. For both C/C++ and Python, we collected data from code repositories ranked 101-200 by GitHub Stars, following the same filtering process as described in Section~\ref{sec-4-setup} (filtering commits modifying single functions, keyword filtering, and LLM verification). We randomly selected 15 optimization-related commits for each language as the pilot study dataset, which accounts for approximately 10\% of the main experiment benchmark (9.93\% for C/C++ with 151 tasks, and 10.00\% for Python with 150 tasks). Since the pilot study dataset does not overlap with the Top 100 code repositories used in the main experiments, the pilot study results are not influenced by the data distribution of the main experiments.

This pilot study employs Semantic Equivalence (SemEqv) as the evaluation criterion, determining whether the optimization results generated by the tool are semantically equivalent to the developers' ground truth patches, as described in Section~\ref{sec-4-setup}. For each optimization task, a method is considered successful if at least one of its generated solutions meets this criterion. The "Successful Optimizations" reported in all tables in this section refer to the total number of successfully optimized tasks out of $15$ pilot study tasks.

When investigating the value of a particular parameter, all other parameters remain consistent with the default settings described in Section~\ref{sec-approach}.

\subsubsection{DBSCAN Clustering Parameters}\label{sec-4-pilot-e1}

This experiment aims to determine the optimal values for the hyperparameters $\varepsilon$ (neighborhood radius, controlling the minimum similarity during clustering) and minimum cluster size (the minimum number of commits required to form a valid cluster) in the DBSCAN algorithm. These two parameters correspond to those mentioned in Section~\ref{sec-3-1-3} and directly affect clustering quality and the effectiveness of the strategy library.

For both C/C++ and Python, we tested the following parameter combinations:
\begin{itemize}
    \item Neighborhood radius $\varepsilon \in \{0.83, 0.85, 0.87, 0.89, 0.90\}$ ($5$ values)
    \item Minimum cluster size $\in \{4, 5, 6\}$ ($3$ values)
\end{itemize}
In total, we conducted $5 \times 3 = 15$ groups of experiments for each language. Table~\ref{tab:4-pilot-e1-combined} presents the DBSCAN parameter sensitivity experimental results for C/C++ and Python. Each row in the tables corresponds to a minimum cluster size value, each column corresponds to a neighborhood radius value, and the cell values indicate the number of successful optimizations under that parameter combination.

\begin{table}[t]
\centering
\caption{DBSCAN Parameter Sensitivity Results.}
\label{tab:4-pilot-e1-combined}
\begin{adjustbox}{max width=\linewidth}
\begin{tabular}{
    >{\centering\arraybackslash}m{1.5cm}
    >{\centering\arraybackslash}m{2cm}
    >{\centering\arraybackslash}m{1.3cm}
    >{\centering\arraybackslash}m{1.3cm}
    >{\centering\arraybackslash}m{1.3cm}
    >{\centering\arraybackslash}m{1.3cm}
    >{\centering\arraybackslash}m{1.3cm}
}
\toprule
\textbf{Language} & \textbf{Min Cluster Size} & \textbf{0.83} & \textbf{0.85} & \textbf{0.87} & \textbf{0.89} & \textbf{0.90} \\
\midrule
 \multirow{3}*{C/C++} & 4 & 3 & 4 & 4 & 5 & 4 \\
  & 5 & 4 & 5 & 5 & 6 & 6 \\
  & 6 & 4 & 5 & 4 & 5 & 4 \\
\midrule
 \multirow{3}*{Python} & 4 & 5 & 4 & 4 & 4 & 5 \\
  & 5 & 4 & 7 & 6 & 5 & 5 \\
  & 6 & 4 & 6 & 4 & 5 & 5 \\
\bottomrule
\end{tabular}
\end{adjustbox}
\end{table}

The experimental results demonstrate that the combination of $\varepsilon=0.89$ and minimum cluster size of $5$ achieves the best performance on C/C++, while the combination of $\varepsilon=0.85$ and minimum cluster size of $5$ achieves the best performance on Python, striking a reasonable balance between filtering noise and retaining effective strategies.

Note that on C/C++, both $\varepsilon=0.89$ with minimum cluster size of $5$ and $\varepsilon=0.90$ with minimum cluster size of $5$ achieve the same number of successful optimizations. However, the former configuration yields a richer strategy library containing $155$ optimization strategies, whereas the latter contains only $140$. Therefore, we ultimately select $\varepsilon=0.89$ with minimum cluster size of $5$ to build a more comprehensive strategy library.

\subsubsection{Number of Rules per Strategy}\label{sec-4-pilot-e2}

This experiment aims to determine the optimal number of commits sampled from each strategy cluster and the number of rules generated per commit. As described in Section~\ref{sec-3-2-1}, sampling multiple commits expands the observable space of code patterns to which the strategy can be applied, improving coverage; while generating multiple independent rules for each commit helps reduce the uncertainty of single-generation and improves rule quality. These two parameters together determine the total number of rules corresponding to each strategy, directly affecting the tool's ability to discover optimization opportunities.

For both C/C++ and Python, we tested $6$ different commit sampling and rule generation configurations. Table~\ref{tab:4-pilot-e2-combined} presents the experimental results for both languages, where the first column indicates the number of commits sampled from each strategy cluster, the second column indicates the number of rules generated per commit, and the third column indicates the number of successful optimizations under that configuration.

\begin{table}[t]
\centering
\caption{Ablation Study Results on Number of Rules.}
\label{tab:4-pilot-e2-combined}
\begin{adjustbox}{max width=.9\linewidth}
\begin{tabular}{
    >{\centering\arraybackslash}m{2.5cm}
    >{\centering\arraybackslash}m{2.5cm}
    >{\centering\arraybackslash}m{3cm}
    >{\centering\arraybackslash}m{3cm}
}
\toprule
\textbf{Commits Sampled} & \textbf{Rules per Commit} & \textbf{C/C++ Successful Optimizations} & \textbf{Python Successful Optimizations} \\
\midrule
1 & 1 & 6.67\% (1/15) & 13.33\% (2/15) \\
1 & 5 & 13.33\% (2/15) & 26.67\% (4/15) \\
3 & 5 & 20.00\% (3/15) & 26.67\% (4/15) \\
5 & 5 & 33.33\% (5/15) & 40.00\% (6/15) \\
8 & 5 & 40.00\% (6/15) & 40.00\% (6/15) \\
10 & 5 & 40.00\% (6/15) & 46.67\% (7/15) \\
\bottomrule
\end{tabular}
\end{adjustbox}
\end{table}

The experimental results indicate that repeatedly generating Semgrep rules and increasing the number of sampled commits both contribute to improving accuracy. However, increasing these parameters also increases the overhead of rule generation and subsequent optimization. Considering both accuracy and computational cost, we consider the configuration with $10$ sampled commits and $5$ rules per commit to be appropriate, and therefore adopt it in the main experiments.

\subsubsection{Top-N Selection for Optimization}\label{sec-4-pilot-e9}

This experiment aims to determine the optimal value for the number of candidates $N$ retained during optimization. As described in Section~\ref{sec-3-3}, Semgrep rules may produce false matches at inappropriate locations, while the number of rules matching the same location can serve as an indicator of the likelihood that the location represents a real optimization opportunity. We rank candidates by the number of matching rules and retain only the top $N$ for optimization. A larger $N$ increases the overhead of rule generation and LLM invocations, while a smaller $N$ may reduce the success rate by missing real optimization opportunities. Therefore, it is necessary to determine an appropriate $N$ value that balances success rate and computational cost.

We tested $N$ values ranging from $1$ to $25$. Specifically, for each optimization task, we generate $N$ optimization candidates based on Semgrep rule scanning results (ranked by the number of matching rules from high to low) and invoke the LLM to generate optimized code for each candidate. We count the number of successful optimizations under different $N$ values.

Table~\ref{tab:4-pilot-e9} illustrates the trend of successful optimization counts under different $N$ values.

\begin{table}[t]
\centering
\caption{Impact of Top-N Selection on Success Rate.}
\label{tab:4-pilot-e9}
\begin{adjustbox}{max width=\linewidth}
\begin{tabular}{
    >{\centering\arraybackslash}m{2cm}
    >{\centering\arraybackslash}m{3cm}
    >{\centering\arraybackslash}m{3cm}
}
\toprule
\textbf{$N$} & \textbf{C/C++ Success Rate} & \textbf{Python Success Rate} \\
\midrule
1 & 23.20\% (35/151) & 13.30\% (20/150) \\
5 & 28.48\% (43/151) & 31.33\% (47/150) \\
10 & 38.41\% (58/151) & 38.67\% (58/150) \\
15 & 39.07\% (59/151) & 40.00\% (60/150) \\
20 & 41.06\% (62/151) & 43.33\% (65/150) \\
23 & 41.72\% (63/151) & 45.33\% (68/150) \\
25 & 42.38\% (64/151) & 46.00\% (69/150) \\
\bottomrule
\end{tabular}
\end{adjustbox}
\end{table}

As shown in Table~\ref{tab:4-pilot-e9}, as $N$ increases, the number of successful optimizations gradually increases. When $N=25$, the number of successful optimizations stabilizes, and the gain from further increasing $N$ is limited. Therefore, we adopt $N=25$ in the main experiments to strike a balance between optimization effectiveness and computational cost.

\subsection{RQ1: Comparison with Baselines}\label{sec-4-rq1}

To evaluate the effectiveness of \toolname in code optimization, we compared it with RAPGen, its enhanced version RAPGen+, and two representative prompt engineering techniques -- retrieval-augmented generation (RAG) and direct prompting -- on the C/C++ benchmark. On the Python benchmark, we evaluated \toolname, RAPGen, and the prompt engineering techniques using DeepSeek-V3 only.
Table~\ref{tab:4-rq1} and Table~\ref{tab:4-rq1-python} present the performance of each tool on the C/C++ benchmark (151 tasks) and the Python benchmark (150 tasks), respectively.

\begin{table}
\centering
\renewcommand{\arraystretch}{1.2}
\caption{The performance comparison of \toolname and baselines on different LLMs on the C/C++ benchmark.}
\label{tab:4-rq1}
\begin{adjustbox}{max width=\linewidth}
\begin{tabular}{
    >{\centering\arraybackslash}m{2.5cm}     
    >{\centering\arraybackslash}m{1.8cm}     
    >{\centering\arraybackslash}m{1.8cm}       
    >{\centering\arraybackslash}m{1.8cm}     
    >{\centering\arraybackslash}m{1.8cm}       
    >{\centering\arraybackslash}m{1.9cm}     
    >{\centering\arraybackslash}m{1.9cm}       
}
\toprule
\multirow{2}{*}{\textbf{Approach}} 
  & \multicolumn{2}{c}{\textbf{DeepSeek-V3}}
  & \multicolumn{2}{c}{\textbf{GPT-4.1}}
  & \multicolumn{2}{c}{\textbf{Gemini-2.5-Pro (on 40 problems)}} \\
\cmidrule(lr){2-3} \cmidrule(lr){4-5} \cmidrule(lr){6-7}
 & EM & SemEqv
 & EM & SemEqv
 & EM & SemEqv \\
\midrule
RAPGen    & 0 (0.00\%)   & 3 (2.00\%)   & 1 (0.70\%)   & 2 (1.30\%) & - & - \\
RAPGen+   & 3 (2.00\%)   & 6 (4.00\%)   & 3 (2.00\%)   & 7 (4.60\%) & - & - \\
Direct    & 2 (1.30\%) & 9 (6.00\%) & 7 (4.60\%) & 13 (8.60\%) & 0 (0.00\%) & 3 (7.50 \%) \\
RAG       & 25 (16.56\%) & 36 (23.80\%) & 17 (11.26\%) & 32 (21.20\%) & 6 (15.00\%) & 9 (22.50\%) \\
\toolname & 42 (27.81\%) & 64 (42.38\%) & 28 (18.54\%) & 44 (29.14\%) & 12 (30.00\%) & 18 (45.00\%) \\
\toolname + RAG & 49 (32.45\%) & 75 (49.67\%) & 37 (24.50\%) & 58 (38.41\%) & 16 (40.00\%) & 21 (52.50\%) \\
\midrule
Clang-tidy (unrelated to LLMs) & 4 (2.65\%) & 4 (2.65\%) & 4 (2.65\%) & 4 (2.65\%) & 3 (7.50\%) & 3 (7.50\%) \\
\bottomrule
\end{tabular}
\end{adjustbox}
\end{table}

\begin{table}
\centering
\renewcommand{\arraystretch}{1.2}
\caption{The performance comparison on the Python benchmark (150 tasks) using DeepSeek-V3.}
\label{tab:4-rq1-python}
\begin{adjustbox}{max width=.7\linewidth}
\begin{tabular}{
    >{\centering\arraybackslash}m{2.5cm}     
    >{\centering\arraybackslash}m{1.8cm}     
    >{\centering\arraybackslash}m{1.8cm}       
}
\toprule
\textbf{Approach} & \textbf{EM} & \textbf{SemEqv} \\
\midrule
RAPGen    & 6 (4.00\%)   & 15 (10.00\%) \\
Direct    & 6 (4.00\%) & 14 (9.33\%) \\
RAG       & 24 (16.00\%) & 43 (28.67\%) \\
\toolname & 38 (25.33\%) & 69 (46.00\%) \\
\toolname + RAG & 50 (33.33\%) & 85 (56.67\%) \\
\bottomrule
\end{tabular}
\end{adjustbox}
\end{table}

\subsubsection{Comparison to RAPGen}

As shown in Table~\ref{tab:4-rq1} and Table~\ref{tab:4-rq1-python}, \toolname achieves notable improvements over RAPGen on all LLMs.
Under the Exact Match metric, RAPGen successfully optimized 0 and 1 cases on the C/C++ benchmark, while \toolname achieved 42 and 28 cases; on the Python benchmark, \toolname achieves 6.33 times more successful optimizations than RAPGen.
Under the Semantic Equivalence metric on the C/C++ benchmark, the number of successful optimizations achieved by \toolname is 21.33 and 22 times higher than that of RAPGen; on the Python benchmark, \toolname achieves 4.60 times more successful optimizations than RAPGen.
All results are based on evaluations with DeepSeek-V3 and GPT-4.1 \cite{liu2024deepseek, openai2025gpt41}.
These results demonstrate the effectiveness of \toolname for code optimization tasks on both C/C++ and Python.

\begin{figure}[t]
    \centering
    \begin{subfigure}{\linewidth}
        \centering
        \includegraphics[width=.6\linewidth]{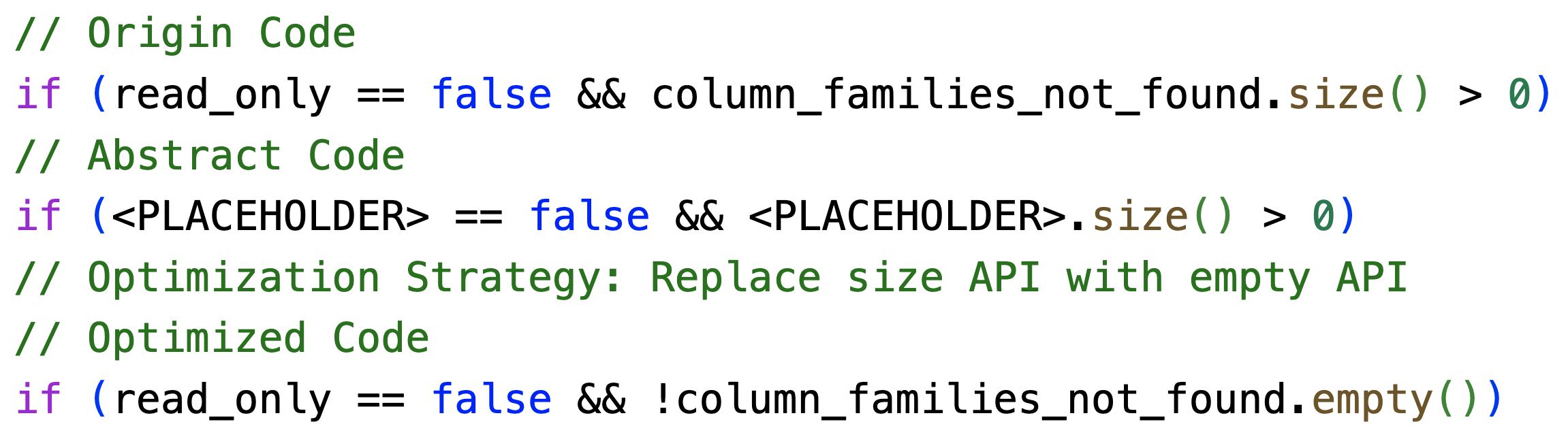}
        \caption{The code requiring optimization.}
        \label{fig:4_rapgen-code_input}
        \vspace{1em}
    \end{subfigure}
    \begin{subfigure}{\linewidth}
        \centering
        \includegraphics[width=.55\linewidth]{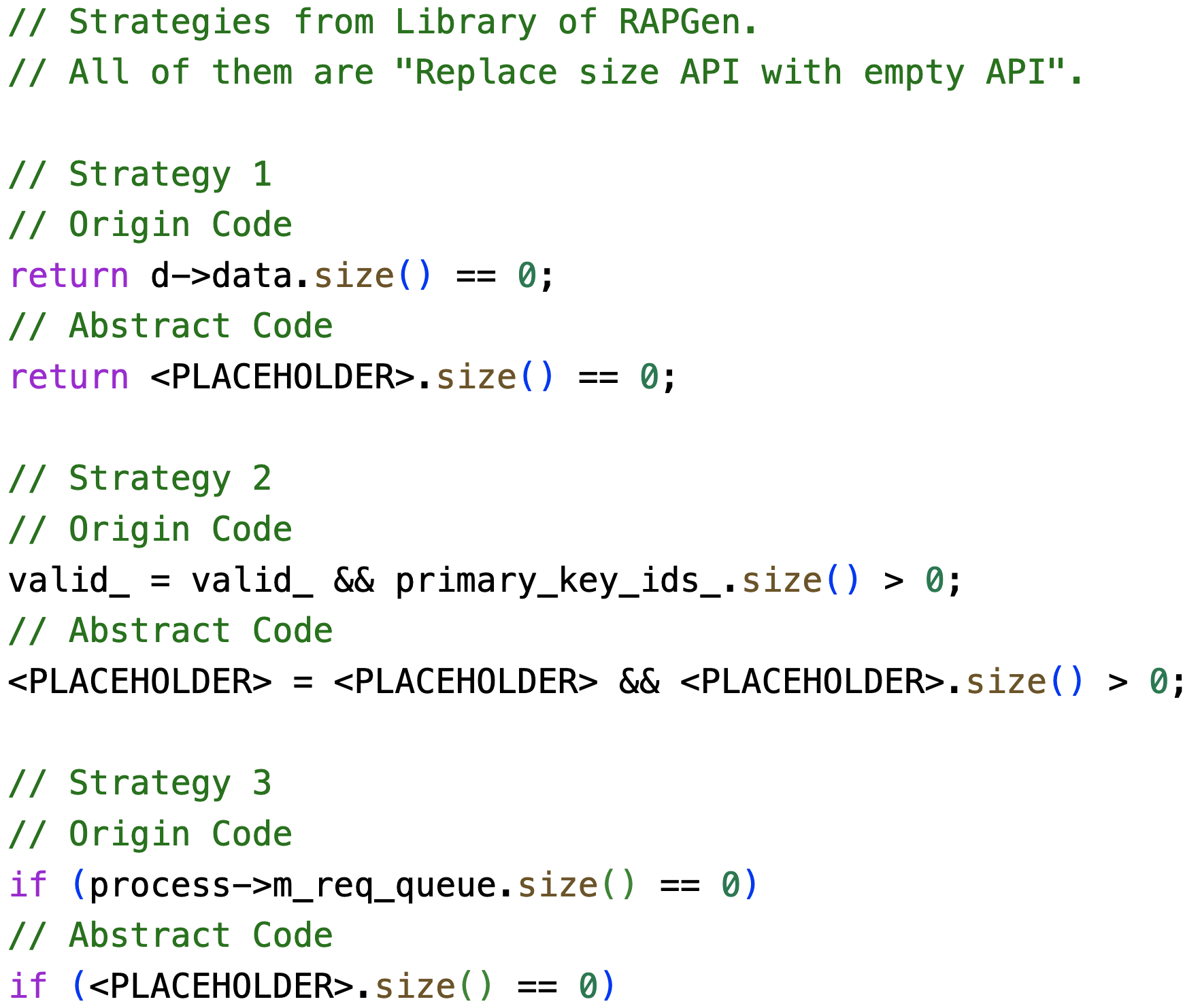}
        \caption{Optimization strategies from RAPGen.}
        \label{fig:4_rapgen-rapgen_strategies}
        \vspace{1em}
    \end{subfigure}
    \begin{subfigure}{\linewidth}
        \centering
        \includegraphics[width=.7\linewidth]{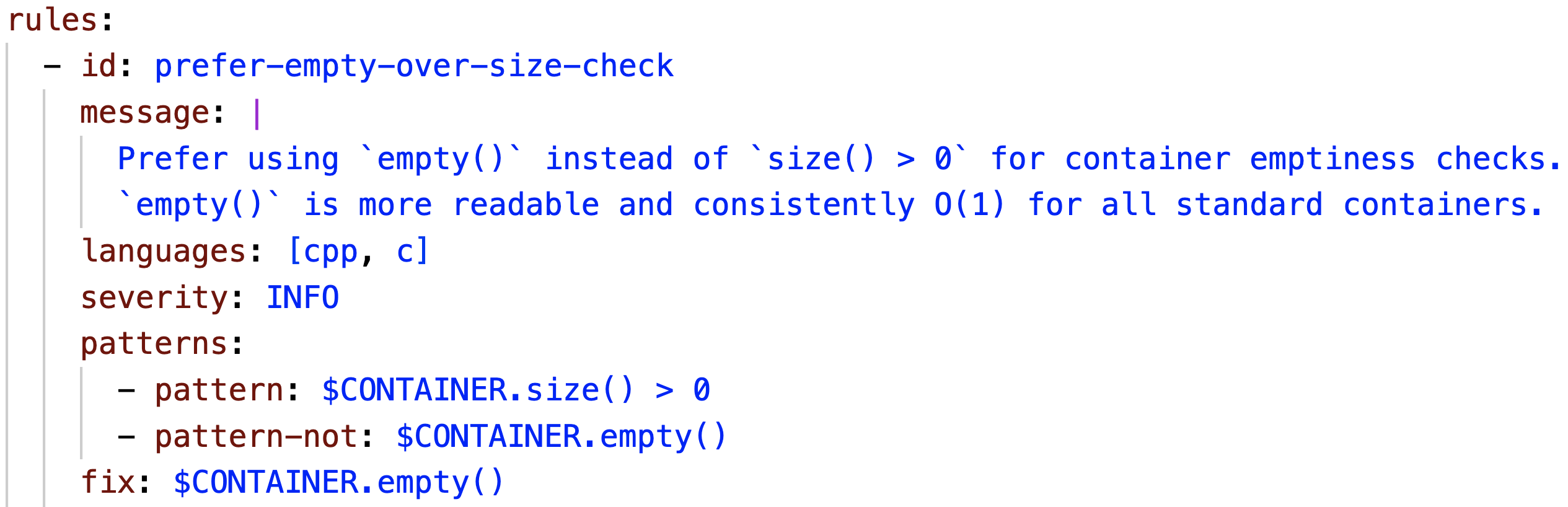}
        \caption{The Semgrep rule from \toolname.}
        \label{fig:4_rapgen-semgrep_rule}
        \vspace{-.5em}
    \end{subfigure}
    \caption{Code optimization example for RQ1.}
    \label{fig:4_rapgen-code_group}
\end{figure}

We studied the reasons behind the optimization failures of RAPGen, which can be summarized into the following three main aspects.

\mainpoint{Limited Strategy:} RAPGen is limited to a single optimization strategy---modifications on individual APIs. However, only about 23 optimization tasks in the benchmark conform to this pattern, which fundamentally restricts the applicability of RAPGen on this benchmark.

\mainpoint{Simplistic Strategy Matching Mechanism:} RAPGen requires that the abstracted code structurally matches the entries in its strategy library exactly, which may result in the omission of effective strategies. Its abstraction approach replaces project-specific identifiers with placeholders, and stores both the source code before optimization and its corresponding abstracted result in the strategy library. A successful match between the code to be optimized and an entry in the strategy library requires that their abstracted representations be exactly identical. As shown in Figure~\ref{fig:4_rapgen-code_group}, although several optimization strategies in the library are consistent with the one depicted in Figure~\ref{fig:4_rapgen-code_input}, the abstracted result of the code to be optimized does not exactly match the abstracted result of any entry in the library, failing to match a strategy. This issue severely limits the optimization capabilities of RAPGen.

\mainpoint{Lack of Location Capability:} RAPGen relies heavily on the input to specify the exact lines of code to be optimized. However, our benchmark does not provide such location information, forcing RAPGen to attempt to match every line in the input code against its strategy library. This process can introduce a large number of incorrect optimization strategies, thereby negatively affecting its overall optimization performance.

\subsubsection{Comparison to RAPGen+}

To evaluate whether RAPGen would perform better without the problem of location, we implemented an enhanced version, RAPGen+, as described in Section~\ref{sec-4-setup}. We explicitly provide the location of the code segments modified in the original commit as additional input to RAPGen+, emulating an ideal condition of precise location. Subsequently, only the code within these segments is used to match optimization strategies, after which similarity is computed to determine the final strategy to apply. This analysis is conducted only on the C/C++ version.

As shown in Table~\ref{tab:4-rq1} and Table~\ref{tab:4-rq1-python}, RAPGen+ demonstrates a slight improvement over RAPGen, with the number of successful optimizations increasing by 2 and 5 cases on the C/C++ benchmark. However, the number of successful optimizations achieved by \toolname is still 14 times (DeepSeek-V3) and 6.29 times (GPT-4.1) greater than that of RAPGen+ on the C/C++ benchmark. This indicates that even with additional correct location information, RAPGen+ still performs significantly worse than \toolname.

\subsubsection{Comparison to Prompt Engineering and RAG}

As shown in Table~\ref{tab:4-rq1} and Table~\ref{tab:4-rq1-python}, the number of successful optimizations achieved by \toolname is much higher than the direct prompt engineering (Direct) and the RAG baseline.
On the C/C++ benchmark under the Exact Match metric, \toolname has 4 to 21 times of the successful optimizations of the Direct baseline, and 1.65 to 2 times of the RAG baseline; under the Semantic Equivalence metric, \toolname has 3.38 to 7.11 times of the Direct baseline, and 1.38 to 2 times of the RAG baseline.
On the Python benchmark under the Exact Match metric, \toolname has 6.33 times of the successful optimizations of the Direct baseline, and 1.58 times of the RAG baseline; under the Semantic Equivalence metric, \toolname has 4.93 times of the Direct baseline, and 1.60 times of the RAG baseline.
These results demonstrate the effectiveness of \toolname in addressing code optimization tasks on both C/C++ and Python.

\subsubsection{Comparison to Clang-tidy}

As shown in Table~\ref{tab:4-rq1}, Clang-tidy only successfully optimized 4 out of 151 C/C++ tasks. In contrast, \toolname successfully optimized 42 to 64 tasks on the same benchmark, which is 10.5 to 16 times of Clang-tidy.

This result indicates that although traditional static analysis tools like Clang-tidy can work well on certain specific patterns, their capabilities heavily depend on predefined check rules. For complex or semantic-aware optimization patterns, traditional static analysis tools struggle to succeed. This result further validates the significant advantage of \toolname in handling complex code optimization tasks.

\subsubsection{Complementarity of \toolname and RAG}

As shown in the \toolname + RAG row of Table~\ref{tab:4-rq1} and Table~\ref{tab:4-rq1-python}, combining the optimization results of both methods increases the number of successful optimizations by 17.19\% to 33.33\% on the C/C++ benchmark and by 23.19\% to 31.58\% on the Python benchmark compared to using \toolname alone. This demonstrates the strong complementarity between \toolname and RAG on both languages.

\subsubsection{Practical Applicability of \toolname}\label{sec-4-applicability}

To evaluate the practicality of \toolname in real-world scenarios, it is necessary to measure the proportion of proposed optimizations that are applicable. An optimization is considered valuable if it preserves the original semantics and achieves a certain degree of performance improvement (e.g., execution speed or resource efficiency), regardless of whether it matches the optimization result in the original submission.

Due to the substantial workload, we conducted a manual evaluation only on all optimization suggestions generated by \toolname when using DeepSeek-V3 on the C/C++ benchmark. Among these, 89.86\% of the optimizations were deemed to meet the criteria as shown in Table~\ref{tab:applicability}. This evaluation result highlights two aspects that supplement the results before:

\begin{table}[t]
\centering
\renewcommand{\arraystretch}{1.2}
\caption{Practical applicability of \toolname.}
\label{tab:applicability}
\begin{adjustbox}{max width=.6\linewidth}
\begin{tabular}{c}
\toprule
Applicable suggestions \\
{89.86\%} \\
\bottomrule
\end{tabular}
\end{adjustbox}
\end{table}

\begin{enumerate}
    \item \textbf{High Practical Applicability of \toolname:} The proportion of applicable suggestions (89.86\%) indicates a low false positive rate, showing that most suggestions produced by \toolname are indeed beneficial to developers in practice, demonstrating the high practical applicability of \toolname in real-world scenarios.
    \item \textbf{Supplementing Evaluation Metrics:} The EM and SemEqv metrics proposed in Section~\ref{sec-4-setup} only count exact matches to the developer's original optimization as successful. This means that if \toolname proposes an effective optimization, but it differs from the developer's patch, it will still be marked as a failure by these metrics. Therefore, these metrics may underestimate the practical effectiveness of \toolname. The high proportion of effective manual results demonstrates that \toolname's real-world usability exceeds what is captured by the EM and SemEqv metrics.
\end{enumerate}

Furthermore, Section~\ref{sec-5-results} further validates this through the test suites provided by real-world projects, showing that the test success rate reaches 72.43\% for C/C++ and 100\% for Python, with 2 optimization proposals having been accepted by project developers, further demonstrating the practical usability of \toolname.

Additionally, Section~\ref{sec-4-rq5} provides a detailed classification analysis of the failure cases. Specifically, it analyzes the cases that failed under the EM and SemEqv metrics (as reported in Section~\ref{sec-4-rq1}) and the cases that failed under the applicability metric (as reported in Section~\ref{sec-4-applicability}), providing insights into the main failure modes and their proportions. 

\begin{rqbox}
\textbf{Answer to RQ1:}
\toolname achieves a significant improvement in producing correct optimization outcomes, reaching 37.50\% to 27x more successful optimizations than all baselines on the C/C++ benchmark and 58.00\% to 5.33x more on the Python benchmark.
Furthermore, \toolname and RAG exhibit strong complementarity.
Besides, \toolname demonstrates high practical applicability, with 89.86\% of its generated optimizations preserving semantics while improving performance.
\toolname demonstrates optimization effectiveness on both C/C++ and Python, indicating that the approach generalizes across programming languages.

\end{rqbox}

\subsection{RQ2: Ablation Study}\label{sec-4-rq2}

We conduct ablation studies on the C/C++ benchmark to validate the effectiveness of each component in \toolname's design.

\begin{table}
\centering
\renewcommand{\arraystretch}{1.2}
\caption{Ablation Study}
\label{tab:4-rq2}
\begin{adjustbox}{max width=\linewidth}
\begin{tabular}{
    >{\arraybackslash}m{4cm}   
    >{\centering\arraybackslash}m{2cm}   
    >{\centering\arraybackslash}m{2cm}   
}
\toprule
\textbf{Approach}
& \textbf{EM}
& \textbf{SemEqv} \\
\midrule
w/o Location   & 36 (23.84\%) & 51 (33.77\%) \\
w/o Optimization Strategy   & 15 (9.93\%)  & 41 (27.15\%) \\
\toolname           & 42 (27.81\%) & 64 (42.38\%) \\
\bottomrule
\end{tabular}
\end{adjustbox}
\end{table}

To verify the effectiveness of the design of \toolname, we conduct an ablation study on the two functionalities associated with Semgrep rules by removing information in the prompt:

\begin{itemize}
    \item \textbf{w/o Location:} Remove the location information from the LLM input prompt and provide only the optimization strategy. The LLM has to determine how to apply the optimization strategy to the whole input function.
    \item \textbf{w/o Optimization Strategy:} Remove the description of the optimization strategy from the LLM input prompt and provide only the location information. The LLM has to determine which strategy to use to optimize the given code segment.
\end{itemize}

We conducted experiments on 151 optimization tasks using DeepSeek-V3. The experimental results are shown in Table~\ref{tab:4-rq2}.
Removing the location information led to a performance drop of 14.29\%–20.31\%, while removing the optimization strategy resulted in a decrease of 35.94\%–64.29\%. These findings demonstrate that both types of information provided by Semgrep rules contribute to the overall performance.

\begin{rqbox}
\textbf{Answer to RQ2:}
In the Semgrep rules generated by \toolname, both location and optimization strategy information significantly contribute to the overall performance.
\end{rqbox}

\subsection{RQ3: Generalization Across Codebases}\label{sec-4-rq3}

This research question focuses on the generalization capability of \toolname across codebases on C/C++. Cross-language generalization has been demonstrated in RQ1 through the Python benchmark.

\begin{table}
\centering
\renewcommand{\arraystretch}{1.2}
\caption{The performance comparison of \toolname{} and prompt engineering techniques on different LLMs.}
\label{tab:4-rq3}
\begin{adjustbox}{max width=\linewidth}
\begin{tabular}{
    >{\centering\arraybackslash}m{1.5cm}     
    >{\centering\arraybackslash}m{1.4cm}     
    >{\centering\arraybackslash}m{1.7cm}     
    >{\centering\arraybackslash}m{1.7cm}       
    >{\centering\arraybackslash}m{1.7cm}     
    >{\centering\arraybackslash}m{1.7cm}       
    >{\centering\arraybackslash}m{1.9cm}     
    >{\centering\arraybackslash}m{1.9cm}       
}
\toprule
\multirow{2}{*}{\textbf{Approach}} 
  & \multirow{2}{*}{\textbf{Mode}} 
  & \multicolumn{2}{c}{\textbf{DeepSeek-V3}}
  & \multicolumn{2}{c}{\textbf{GPT-4.1}}
  & \multicolumn{2}{c}{\textbf{Gemini-2.5-Pro (on 40 problems)}} \\
\cmidrule(lr){3-4} \cmidrule(lr){5-6} \cmidrule(lr){7-8}
 & & EM & SemEqv
 & EM & SemEqv
 & EM & SemEqv \\
\midrule
\multirow{2}{*}{RAG}
    & Standard   & 25 (16.56\%) & 36 (23.80\%) & 17 (11.26\%) & 32 (21.20\%) & 6 (15.00\%) & 9 (22.50\%) \\
    & Degraded   & 6 (3.97\%) & 10 (6.62\%) & 5 (3.31\%) & 15 (9.93\%) & 1 (2.50\%) & 5 (12.50\%) \\
\midrule
\multirow{2}{*}{\toolname}
    & Standard   & 42 (27.81\%) & 64 (42.38\%) & 28 (18.54\%) & 44 (29.14\%) & 12 (30.00\%) & 18 (45.00\%) \\
    & Degraded   & 42 (27.81\%) & 63 (41.72\%) & 23 (15.23\%) & 42 (27.81\%) & 9 (22.50\%) & 16 (40.00\%) \\
\bottomrule
\end{tabular}
\end{adjustbox}
\end{table}

\begin{figure}[t]
    \centering
    \begin{subfigure}{\linewidth}
        \centering
        \includegraphics[width=.68\linewidth]{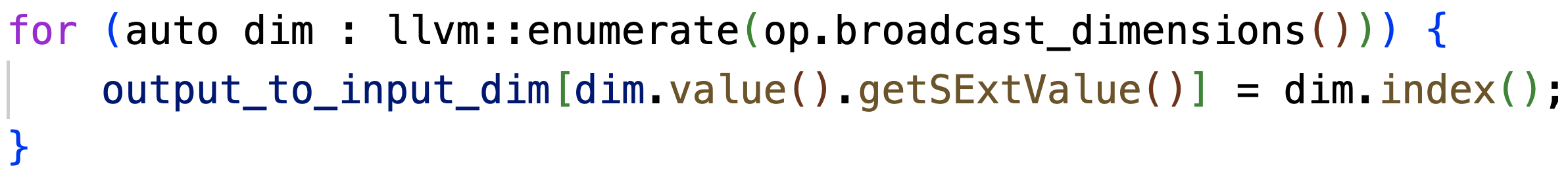}
        \caption{The code segment requiring optimization.}
        \label{fig:4_rag-code_input}
        \vspace{1em}
    \end{subfigure}
    \begin{subfigure}{\linewidth}
        \centering
        \includegraphics[width=.7\linewidth]{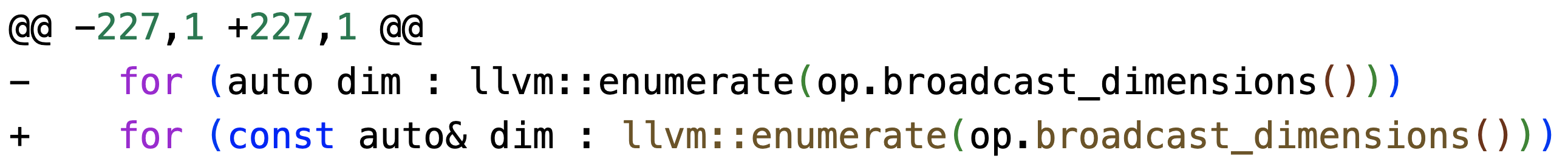}
        \caption{The correct optimization result.}
        \label{fig:4_rag-code_diff}
        \vspace{1em}
    \end{subfigure}
    \begin{subfigure}{\linewidth}
        \centering
        \includegraphics[width=.75\linewidth]{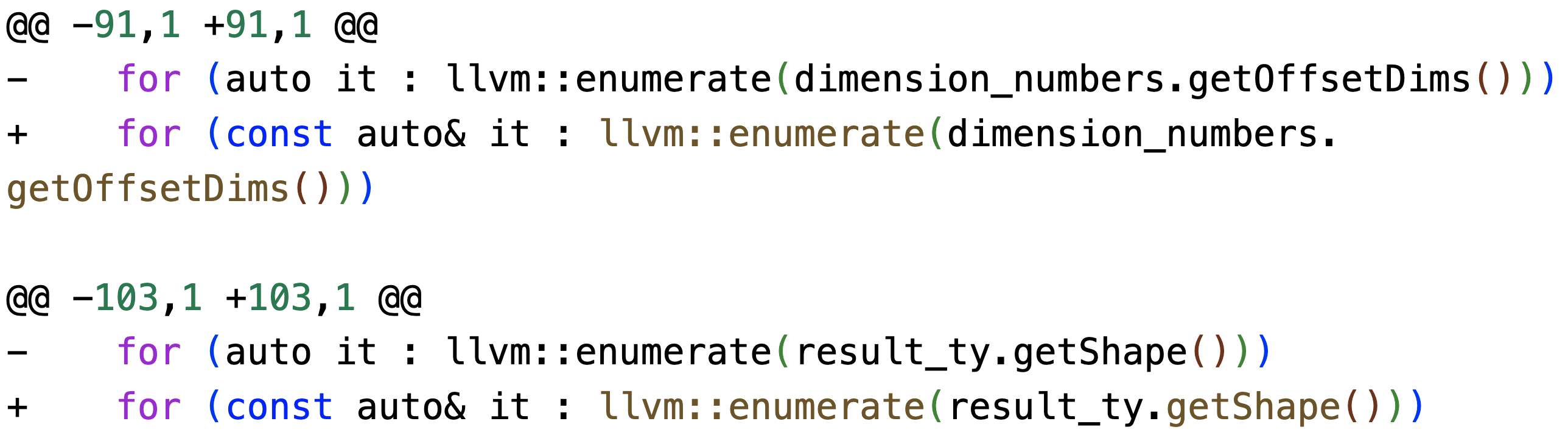}
        \caption{The first code modification reference used by RAG.}
        \label{fig:4_rag-similar_diff_1}
        \vspace{1em}
    \end{subfigure}
    \begin{subfigure}{\linewidth}
        \centering
        \includegraphics[width=.55\linewidth]{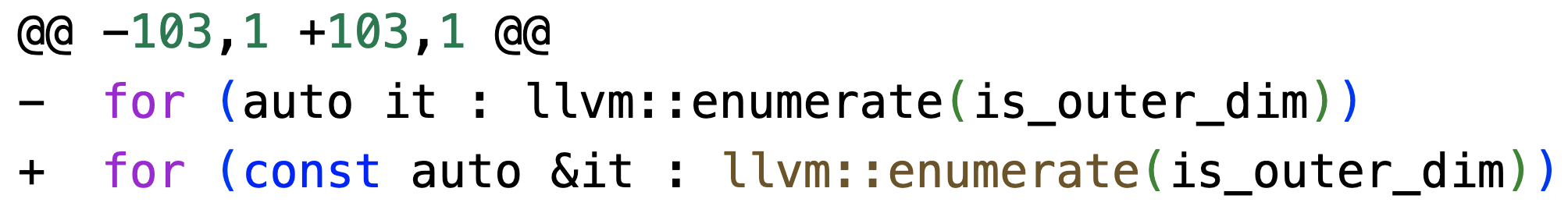}
        \caption{The second code modification reference used by RAG.}
        \label{fig:4_rag-similar_diff_2}
        \vspace{-.5em}
    \end{subfigure}
    \caption{Code optimization example for RQ3.}
    \label{fig:4_rag-code_group}
\end{figure}

We observed that, for some tasks where RAG achieved successful optimizations, it referenced other commits from the same codebase. In certain cases, the optimization strategies in the referenced commits were highly consistent with those required by the current task, thus providing substantial assistance.

An example is presented in Figure~\ref{fig:4_rag-code_group}. Figure~\ref{fig:4_rag-code_input} shows the code snippet to be optimized, while Figure~\ref{fig:4_rag-code_diff} illustrates the correct optimization approach, in which the loop variable in the for statement is modified with both const and reference qualifiers.
Figures~\ref{fig:4_rag-similar_diff_1} and \ref{fig:4_rag-similar_diff_2} present the top-1 and top-3 most similar examples retrieved by the RAG method as references for code optimization. All three code optimization commits originate from the TensorFlow~\cite{tensorflow} codebase and were submitted by the same developer on the same day, with largely consistent optimization strategies.

To further investigate the dependence of \toolname and RAG on information from the same codebase, we implemented two modes for each method: Standard Mode and Degraded Mode.

\begin{itemize}
    \item \textbf{Standard Mode:} Correspond to the standard implementation used in RQ1. In this mode, we exclude the exact match of the commit or code from the strategy library and the RAG knowledge base to avoid data leakage, but there may still be similar commits in the same codebase, such as the example in Figure~\ref{fig:4_rag-code_group}.
    \item \textbf{Degraded Mode:} Corresponds to a reduced implementation that excludes the knowledge in the same codebase. During optimization, all entries in the knowledge base corresponding to historical commits from the same codebase are excluded, while all other entries remain unchanged.
\end{itemize}

As shown in Table~\ref{tab:4-rq3}, when switching from Standard Mode to Degraded Mode, the number of successful optimizations achieved by \toolname decreases by only 0\% to 25\%, while RAG experiences a substantially larger reduction of 44.44\% to 83.33\%.
Under both metrics in Degraded Mode, \toolname achieves 2.8 to 9 times as many successful optimizations as RAG.
This demonstrates that the majority of RAG's successful optimizations depend on historical commits from the same codebase, whereas \toolname shows significantly less reliance, indicating that \toolname has stronger generalization capability across different codebases.

\begin{rqbox}
\textbf{Answer to RQ3:}
RAG demonstrates a significantly higher dependence on information from other commits within the same codebase compared to \toolname, indicating that \toolname has stronger generalization capabilities across codebases.
\end{rqbox}

\subsection{RQ4: Taxonomy Analysis of Optimization Strategies}\label{sec-4-rq4}

This research question focuses on the scope and diversity of optimization strategies covered by the \toolname strategy library, as well as the optimization effectiveness on each type of benchmark. This analysis is conducted only on the C/C++ version.

To systematically understand the scope and diversity of optimization strategies covered by the \toolname strategy library, we construct a taxonomy with 10 categories, and classify both the 155 strategies in the library and the 151 benchmarks in Section~\ref{sec-empirical-eval}. Table~\ref{tab:taxonomy-combined} presents the distribution of these 10 categories, along with \toolname's (using DeepSeek-V3) Exact Match (EM) and Semantic Equivalence (SemEqv) success rates for each category.

\begin{table}[t]
  \caption{\toolname Optimization Strategy Taxonomy Analysis.} \label{tab:taxonomy-combined}
  \centering
  \small
  \begin{tabular}{
    l
    r@{\hspace{0.2em}}r
    r@{\hspace{0.2em}}r
    r@{\hspace{0.2em}}r
    r@{\hspace{0.2em}}r
  }
    \toprule
    \multirow{2}{*}{\textbf{Category}} & \multicolumn{2}{c}{\textbf{\#Benchmarks}} & \multicolumn{2}{c}{\textbf{\#Strategies}} & \multicolumn{2}{c}{\textbf{EM Success}} & \multicolumn{2}{c}{\textbf{SemEqv Success}} \\
    \cmidrule(lr){2-3} \cmidrule(lr){4-5} \cmidrule(lr){6-7} \cmidrule(lr){8-9}
    & Count & (\%) & Count & (\%) & Count & (\%) & Count & (\%) \\
    \midrule
    Memory Pre-allocation & 18 & (11.92\%) & 10 & (6.45\%) & 3/18 & (16.67\%) & 7/18 & (38.89\%) \\
    Avoiding Unnecessary Copies & 50 & (33.11\%) & 13 & (8.39\%) & 26/50 & (52.00\%) & 30/50 & (60.00\%) \\
    Caching \& Redundant Computation Elimination & 17 & (11.26\%) & 20 & (12.90\%) & 1/17 & (5.88\%) & 9/17 & (52.94\%) \\
    Loop Structure Optimization & 6 & (3.97\%) & 6 & (3.87\%) & 2/6 & (33.33\%) & 2/6 & (33.33\%) \\
    Data Structure \& Algorithm Replacement & 6 & (3.97\%) & 15 & (9.68\%) & 0/6 & (0.00\%) & 2/6 & (33.33\%) \\
    Efficient API \& Library Usage & 26 & (17.22\%) & 18 & (11.61\%) & 7/26 & (26.92\%) & 7/26 & (26.92\%) \\
    Conditional \& Branch Optimization & 18 & (11.92\%) & 18 & (11.61\%) & 1/18 & (5.56\%) & 5/18 & (27.78\%) \\
    Parallelization \& Concurrency & 2 & (1.32\%) & 24 & (15.48\%) & 1/2 & (50.00\%) & 1/2 & (50.00\%) \\
    Compiler Hints \& Low-Level Optimization & 6 & (3.97\%) & 10 & (6.45\%) & 1/6 & (16.67\%) & 1/6 & (16.67\%) \\
    I/O \& System Resource Optimization & 2 & (1.32\%) & 21 & (13.55\%) & 0/2 & (0.00\%) & 0/2 & (0.00\%) \\
    \midrule
    \textbf{Total} & \textbf{151} & \textbf{(100.00\%)} & \textbf{155} & \textbf{(100.00\%)} & \textbf{42}/151 & \textbf{(27.81\%)} & \textbf{64}/151 & \textbf{(42.38\%)} \\
    \bottomrule
  \end{tabular}
\end{table}

\toolname's strategy library covers diverse optimization types, with all 10 categories present in both the benchmarks and the strategy library. Among them, Avoiding Unnecessary Copies is the most common optimization type (50 benchmarks, 33.11\%), and also the category where \toolname performs best, with an EM success rate of 52.00\% and SemEqv success rate of 60.00\%. In contrast, Data Structure \& Algorithm Replacement and I/O \& System Resource Optimization are more challenging for \toolname due to the high difficulty of these optimization tasks.

\begin{rqbox}
\textbf{Answer to RQ4:} \toolname's strategy library covers diverse optimization types. Avoiding Unnecessary Copies is the most common and best-performing category. Categories with higher task difficulty have lower optimization success rates for \toolname.
\end{rqbox}

\subsection{RQ5: Failure Case Analysis}\label{sec-4-rq5}

This research question aims to analyze the main failure modes of \toolname and their proportions. This analysis is conducted only on the C/C++ version. Specifically, we perform two types of failure analysis: (1) analysis based on EM and SemEqv metrics for the cases that did not pass these metrics (as reported in Section~\ref{sec-4-rq1}), and (2) analysis based on the applicability metric for the cases that passed the first evaluation but failed the second (as reported in Section~\ref{sec-4-applicability}).

Since the optimization tasks in the benchmark come from existing optimization tasks on GitHub, we treat the modifications in the original commits as the ground truth, expecting to reproduce the optimization. Besides, we also acknowledge that \toolname may provide correct optimization solutions that differ from the developer's but are actually acceptable; these are not counted as success in this analysis. See Section~\ref{sec-4-applicability} for detailed discussion.

\vspace{1em}
\noindent\textbf{Part 1: Failure Analysis Based on EM and SemEqv Metrics.}

Before categorizing the failure cases, we need to first define how to determine whether a Semgrep rule's scanning result on a benchmark is "correct". The determination includes two dimensions:
\begin{itemize}
    \item \textbf{Correct Location:} The code fragment matched by Semgrep overlaps with the ground truth modification range, with an overlap ratio of at least 50\% (i.e., the matched fragment covers at least 50\% of the ground truth modification range).
    \item \textbf{Correct Strategy:} The optimization strategy is consistent with the strategy actually used in the ground truth.
\end{itemize}
Only when both correct location and correct strategy are satisfied, the scanning result is considered correct.

Based on this, in Section~\ref{sec-4-rq1}, using the results optimized by DeepSeek-V3, we conduct a systematic analysis on the $109$ C/C++ failure cases that did not pass the EM metric. We categorize the failure reasons according to the pipeline stages of \toolname into $4$ categories (with $6$ subcategories), as shown in Table~\ref{tab-failure-analysis}.

\begin{table}[htbp]
\centering
\caption{Failure Case Analysis for EM and SemEqv Metrics}
\label{tab-failure-analysis}
\begin{adjustbox}{max width=\textwidth}
\begin{tabular}{
    >{\centering\arraybackslash}m{3.8cm}
    >{\centering\arraybackslash}m{5.4cm}
    >{\centering\arraybackslash}m{1.3cm}
    >{\centering\arraybackslash}m{1.6cm}
}
\toprule
\textbf{Category} & \textbf{Subcategory} & \textbf{Count} & \textbf{Percentage} \\
\midrule
Strategy Library Coverage & — & 4 & 3.67\% \\
\midrule
Semgrep Rule Miss & Semgrep Expressiveness Limitation & 1 & 0.90\% \\
 & LLM Rule Coverage Insufficient & 21 & 19.27\% \\
 & \textbf{Subtotal} & \textbf{22} & \textbf{20.18\%} \\
\midrule
LLM Optimization Failure & Correct Strategy Not in Top-25 & 36 & 33.03\% \\
 & Correct Strategy in Top-25 & 23 & 21.10\% \\
 & \textbf{Subtotal} & \textbf{59} & \textbf{54.13\%} \\
\midrule
Semantically Equivalent & — & 24 & 22.02\% \\
\bottomrule
\end{tabular}
\end{adjustbox}
\end{table}

The categories are described as follows:

\begin{itemize}
    \item \textbf{Strategy Library Coverage (4 cases, 3.67\%):} The strategy library does not contain optimization strategies semantically related to the target optimization task.

    \item \textbf{Semgrep Rule Miss (22 cases, 20.18\%):} The strategy library contains semantically related optimization strategies, but the corresponding Semgrep rules fail to match relevant code fragments in the target function. These 22 cases can be further divided into two subcategories. One subcategory (1 case) is because the optimization pattern itself exceeds the expressiveness of Semgrep and cannot be expressed using Semgrep syntax-level pattern matching. The other subcategory (21 cases) is because although the rules generated by LLM can theoretically express the optimization pattern, the rule patterns are not general enough to cover the specific coding style in the target code.

    \item \textbf{LLM Optimization Failure (59 cases, 54.13\%):} The Semgrep rules corresponding to the correct strategy matched relevant code fragments in the target function, but LLM failed to generate optimization results consistent with the developer's patch. These 59 cases can be further divided into two subcategories. In one subcategory (36 cases), the code fragments corresponding to the correct strategy failed to enter top-25 after ranking, so LLM never received the correct optimization hint. In the other subcategory (23 cases), the correct strategy fragments are already in top-25, LLM received the correct optimization hint but still failed to generate correct optimization.

    \item \textbf{Semantically Equivalent (24 cases, 22.02\%):} The Semgrep rules corresponding to the correct strategy matched relevant code fragments in the target function, and LLM successfully generated optimization results. However, the implementation differs from the developer's, resulting in EM failure. Nevertheless, these optimization results are semantically equivalent to the developer's patch and can be considered successful under the more lenient SemEqv evaluation metric.
\end{itemize}

\vspace{1em}
\noindent\textbf{Part 2: Failure Analysis Based on Applicability Metric.}

To further understand the $89.86\%$ applicability rate reported in Section~\ref{sec-4-applicability}, we conducted an in-depth analysis of the $237$ cases that passed the first evaluation but failed the second evaluation. We categorized the failure reasons into $3$ major categories (with $8$ subcategories), as shown in Table~\ref{tab:applicability-failure-analysis}.

\begin{table}[t]
\centering
\caption{Failure Case Analysis for 89.86\% Applicability Rate.}
\label{tab:applicability-failure-analysis}
\begin{adjustbox}{max width=\linewidth}
\begin{tabular}{
    >{\centering\arraybackslash}m{2.8cm}
    >{\centering\arraybackslash}m{6.6cm}
    >{\centering\arraybackslash}m{1.0cm}
    >{\centering\arraybackslash}m{1.3cm}
}
\toprule
\textbf{Category} & \textbf{Subcategory} & \textbf{Count} & \textbf{Percentage} \\
\midrule
Semantic Issues & Functional behavior change & 43 & 18.14\% \\
 & Boundary condition problems & 50 & 21.10\% \\
 & Potential bugs/undefined behavior & 64 & 27.00\% \\
 & \textbf{Subtotal} & \textbf{157} & \textbf{66.24\%} \\
\midrule
Compilation Issues & Syntax/type errors & 21 & 8.86\% \\
 & Missing dependencies & 3 & 1.27\% \\
 & Platform/compiler compatibility issues & 7 & 2.95\% \\
 & \textbf{Subtotal} & \textbf{31} & \textbf{13.08\%} \\
\midrule
Performance Issues & Overlapping with compiler optimizations & 21 & 8.86\% \\
 & Uncertain performance improvement & 28 & 11.81\% \\
 & \textbf{Subtotal} & \textbf{49} & \textbf{20.68\%} \\
\bottomrule
\end{tabular}
\end{adjustbox}
\end{table}

We manually analyzed the 237 inconsistent cases and categorized them into three main categories, as shown in Table~\ref{tab:applicability-failure-analysis}.

\begin{itemize}
    \item \textbf{Semantic Issues (157 cases, 66.24\%):} The primary cause of inconsistency, including potential bugs/undefined behavior (64 cases, 27.00\%) and boundary condition problems (50 cases, 21.10\%).

    \item \textbf{Compilation Issues (31 cases, 13.08\%):} Technical feasibility issues such as syntax/type errors, missing dependencies, or platform-specific extensions, indicating the optimization suggestions are generally reliable at the technical level.

    \item \textbf{Performance Issues (49 cases, 20.68\%):} Cases where performance improvement is uncertain, including overlaps with compiler optimizations (21 cases, 8.86\%) and uncertain performance improvement (28 cases, 11.81\%).
\end{itemize}

Overall, this analysis further validates the reliability of our reported 89.86\% applicability rate: the vast majority of inconsistencies are caused by issues inherent in the code itself (such as semantic inconsistency or compilation errors), rather than by randomness in the evaluation methodology.

\begin{rqbox}
\textbf{Answer to RQ5:}
For the EM/SemEqv metrics, the failure cases involve strategy library coverage issues, Semgrep rule miss, LLM optimization failure, and semantic equivalence cases. For the applicability metric, the failure cases involve semantic issues, compilation issues, and performance issues.
\end{rqbox}

\subsection{RQ6: Effectiveness of Semgrep Code Location and Ranking Mechanisms}\label{sec-4-rq6}

This research question evaluates the effectiveness of the Semgrep code location and ranking mechanisms in \toolname. This analysis is conducted only on the C/C++ version.

Since the optimization tasks in the benchmark come from existing optimization tasks on GitHub, we treat the modifications in the original commits as the ground truth (see Section~\ref{sec-4-rq5} for details).

Regarding how to determine whether a scanning result is correct, we follow the definitions in Section~\ref{sec-4-rq5}. Precision is defined as the ratio of matches that satisfy both correct location and correct strategy to the total number of matches. Recall is defined as the ratio of benchmarks with at least one match to the total number of benchmarks. F1 score is the harmonic mean of precision and recall.

To evaluate the effectiveness of the Semgrep rule location step in \toolname, we calculate the Precision, Recall, and F1 score for the location recognition stage. We adopt a two-layer metric system, where the first layer measures the accuracy of Semgrep's original scanning, and the second layer measures the final effect after top-N filtering.

\begin{table}[t]
\centering
\caption{Precision, Recall, and F1 of Semgrep code location and ranking mechanisms.}
\label{tab:4-rq6-precision-recall}
\begin{adjustbox}{max width=\linewidth}
\begin{tabular}{
    >{\centering\arraybackslash}m{4cm}
    >{\centering\arraybackslash}m{2.5cm}
    >{\centering\arraybackslash}m{2.5cm}
    >{\centering\arraybackslash}m{2.5cm}
}
\toprule
\textbf{Filtering Condition} & \textbf{Precision} & \textbf{Recall} & \textbf{F1} \\
\midrule
Semgrep Raw Scan & 1.56\% & 82.78\% & 3.06\% \\
After top-N=25 & 5.13\% & 58.94\% & 9.44\% \\
\bottomrule
\end{tabular}
\end{adjustbox}
\end{table}

Table~\ref{tab:4-rq6-precision-recall} presents the detailed experimental results. In the raw scanning stage, Semgrep rules have low precision (1.56\%) but high recall (82.78\%), indicating that Semgrep rules have high sensitivity in locating optimization opportunities and can capture most code locations that need optimization. After top-N=25 filtering, precision increases from 1.56\% to 5.13\%, and F1 score increases from 3.06\% to 9.44\%, indicating that the ranking and filtering mechanisms can effectively improve the ability to identify code that truly needs optimization.

It should be noted that the criterion for determining "correct" here is that the optimization strategy matches the ground truth. However, there may be other correct and acceptable optimization solutions that differ from the ground truth. For details, refer to Section~\ref{sec-4-applicability}. Based on the high proportion of applicable suggestions (89.86\%) in this analysis, it can be concluded that \toolname does not waste many LLM calls.

\begin{rqbox}
\textbf{Answer to RQ6:}
The Semgrep raw scanning in the location recognition stage has low precision (1.56\%) but high recall (82.78\%), indicating that Semgrep rules have high sensitivity in locating optimization opportunities and can capture most code locations that need optimization. After top-N=25 filtering, precision increases to 5.13\% and F1 score increases to 9.44\%, indicating that the ranking and filtering mechanisms can effectively improve the ability to identify code that truly needs optimization.

\end{rqbox}


\section{In-the-Wild Evaluation}\label{sec-5}

In this section, we evaluate the real-world practicality of \toolname by identifying and optimizing hotspot functions in five large-scale C/C++ projects and five large-scale Python projects, then measuring performance improvements using their comprehensive performance test suites. This dual-language evaluation demonstrates that \toolname effectively optimizes real-world projects across different programming languages.


\begin{table}
\centering
\renewcommand{\arraystretch}{1.2}
\caption{In-the-wild evaluation result on C/C++ and Python projects.}
\label{tab:5}
\vspace{-.5em}
\begin{adjustbox}{max width=\linewidth}
\begin{tabular}{
    >{\centering\arraybackslash}m{1.2cm}   
    >{\centering\arraybackslash}m{2cm}   
    >{\centering\arraybackslash}m{1.5cm}   
    >{\centering\arraybackslash}m{1.5cm} 
    >{\centering\arraybackslash}m{1.5cm}   
    >{\centering\arraybackslash}m{1.5cm}   
}
\toprule
\textbf{Lang} & \textbf{Repo}
  & \multicolumn{2}{c}{\textbf{Perf Improvement $\uparrow$}} 
  & \multicolumn{2}{c}{\textbf{\# Test Cases}} \\
\cmidrule(lr){3-4} \cmidrule(lr){5-6}
  &  & Max & Avg
  & $\uparrow \ \geq 5\%$ & $\uparrow \ \geq 10\%$ \\
\midrule
\multirow{5}{*}{C/C++}
  & RocksDB     & 5.04\%   & 2.41\%  & 1/2  & 0/2 \\
  & Redis       & 6.14\%   & 1.68\%  & 3/21 & 0/21 \\
  & gRPC        & 35.48\%  & 3.10\%  & 2/16 & 1/16 \\
  & LevelDB     & 218.07\% & 10.40\% & 2/22 & 1/22 \\
  & spdlog      & 20.00\%  & 3.27\%  & 6/45 & 4/45 \\
\midrule
\multirow{5}{*}{Python}
  & click       & 479.90\% & 143.25\% & 11/11 & 11/11 \\
  & flask       & 66.07\%  & 56.67\%  & 8/8   & 8/8 \\
  & jinja      & 71.43\%  & 48.60\%  & 8/8   & 7/8 \\
  & requests    & 85.45\%  & 40.19\%  & 9/9   & 8/9 \\
  & scrapy      & 61.77\%  & 50.26\%  & 4/4   & 4/4 \\
\bottomrule
\end{tabular}
\end{adjustbox}
\end{table}

\subsection{Experimental Setup}

\subsubsection{Benchmark}

We collected five large-scale C/C++ projects and five large-scale Python projects for evaluation.

\textbf{C/C++ Projects:} We collected RocksDB~\cite{rocksdb}, Redis~\cite{redis}, gRPC~\cite{grpc}, LevelDB~\cite{leveldb}, and spdlog~\cite{spdlog}. These projects are widely used in the industry and span a variety of domains, including storage engines, in-memory databases, distributed communication, remote procedure calls, and logging systems.

\textbf{Python Projects:} We collected click~\cite{click}, flask~\cite{flask}, jinja~\cite{jinja}, requests~\cite{requests}, and scrapy~\cite{scrapy}. These projects cover command-line interfaces, web frameworks, template engines, HTTP libraries, and web crawling frameworks, and are widely influential in the Python ecosystem.

Additionally, each of these projects provides a comprehensive performance test suite, which can be used to quantitatively measure the degree of optimization.

For each project, we cloned the latest available version and wrote a script to compile the code (for C/C++) or install the package (for Python) and run the built-in unit and performance tests. All C/C++ projects are built in release mode with compiler optimizations enabled.
The performance testing results of these 10 projects contain between 2 and 45 distinct test cases for C/C++ projects and between 4 and 11 test cases for Python projects, each test case typically reflecting the performance of a different functionality within the project. We designed a consistent method for extracting the relevant data from each project.

\subsubsection{Metrics}

For each project, we attempt to apply the optimization, compile the code (for C/C++) or install the package (for Python), and then execute the unit tests on the optimized code. If the tests are passed, we conduct the performance tests $6$ times each on both the original and optimized versions of the project. To avoid issues such as cold starts, we discard the results from the first run. For C/C++ projects, the default compiler optimization level defined in the build file of the original project is used. We then report the average performance improvement in the remaining $5$ runs for each test case, which is standard in existing studies~\cite{gao2024search, chen2016robust, liu2024minotaur, wen2025unveiling, fraile2025measuring}.

Depending on the implementation of the test case, the performance may be reported as a number where higher values indicate better performance (e.g., processed items per second) or lower values indicate better performance (e.g., execution time). We unified the performance improvement in both types as the improvement of speed (i.e., 100\% improvement if the number increases from 10 to 20 in the former type, or decreases from 20 to 10 in the latter type). Formally, let $x$ and $y$ denote the performance results for a given test case before and after optimization. The improvement ratio is calculated as $\frac{y - x}{x}$ for the former type or $\frac{x - y}{y}$ for the latter type.

To evaluate the effectiveness of the optimization, we employed the following two metrics.

\begin{itemize}
    \item \textbf{Perf Improvement \boldmath{$\uparrow$}:} The maximum and average improvement ratio of all test cases.
    \item \textbf{\# Test Cases where \boldmath{$\uparrow \; \geq n\%$}:} The number of test cases among all test cases where the improvement ratio exceeds n\%.
\end{itemize}

Furthermore, to investigate whether \toolname identifies new optimizations that are not easy for existing tools, we use an industrial rule-based optimization tool to reproduce the optimizations found by \toolname on C/C++ projects. If the optimization cannot be reproduced, it shows that \toolname is at least complementary to the rule-based tool. We choose JetBrains CLion~\cite{clion} because it is widely adopted in industry and integrates static analysis and refactoring capabilities, making it a representative rule-based tool for automated code optimization in C/C++ development. For each optimization produced by \toolname, we open the corresponding files and locations in CLion and examine the IDE's code inspections, quick fixes, and refactorings. We accept suggestions that the IDE considers safe and automatically applicable. We consider an optimization reproduced if the tool performs a semantically equivalent transformation and the transformed achieves comparable performance improvements to that of \toolname on the same test cases.




\subsection{Optimization Process of \toolname}

For each project, we first identify hotspot functions through profiling, then generate potential optimizations for each function, and finally combine optimizations across all hotspot functions.

\subsubsection{Identify Hotspot Functions}

Since a performance test typically only covers a small set of functions that need to be optimized, we first identify the hotspot functions for each test case. We profile the execution of the performance test: for C/C++ projects, we use \code{perf}~\cite{perf} and \code{gcov}~\cite{gcov}; for Python projects, we use cProfile~\cite{python_profilers} and line\_profiler~\cite{lineprofiler}. Based on the profiling result, we then identify all functions whose execution time accounted for more than 0.1\% of the total runtime as hotspot functions in the performance tests. Only these functions are targets for subsequent optimization.

\subsubsection{Generate and Combine Optimization Results}

For each project, we first apply \toolname to optimize all identified hotspot functions.  

Next, we evaluate all optimization results according to the following criteria: an optimization is considered effective if it achieves a performance improvement exceeding 5\% on at least one test case while causing no more than 2\% performance degradation on any remaining test cases. This approach ensures that selected optimizations provide substantial performance benefits without introducing significant regressions elsewhere.

We then automatically compose the optimization results across different functions through the following process. For each function, if multiple optimization variants satisfy our effectiveness criteria mentioned before, we select the variant with the highest total improvement score -- computed by adding the performance improvement percentages across all test cases. If no variants meet the criteria, the original implementation is preserved for that function. Finally, we assemble the selected optimizations for all hotspot functions to generate the final optimized version of each project.

\subsubsection{Implementation Details}

Unlike the empirical evaluation in Section~\ref{sec-empirical-eval}, this experiment relies on the performance testing results. Therefore, it is crucial to maintain steady performance measurement throughout the experiment.
The experiments were conducted on a Linux server equipped with four Intel Xeon Platinum 8270 CPUs (104 cores @ 2.70 GHz) and 256 GB of RAM.
To minimize the impact of fluctuations on the performance testing results, we ensured that no other resource-intensive programs were running on the server when executing the performance tests.

For all optimization tasks in this section, we used DeepSeek-V3 (version 0324)~\cite{liu2024deepseek} to generate optimization results.

Besides, \toolname is capable of optimizing any user-specified function. We focus on hotspot functions because optimizing these functions is likely to be a more cost-effective choice. However, this does not imply that the operation of \toolname depends on performance test.

\subsection{Optimization Result}\label{sec-5-results}

\textbf{C/C++ Version:} As shown in Table~\ref{tab:5}, the maximum improvement on each C/C++ project ranges from 5.04\% to 218.07\%, with an average improvement between 1.68\% and 10.40\%. 
Each C/C++ project contains 2 to 45 test cases, among which 1 to 6 test cases exhibit improvements greater than 5\%, and 0 to 4 test cases show improvements exceeding 10\%.

We also examined whether CLion could reproduce the optimizations produced by \toolname on C/C++ projects. Across all five C/C++ projects, CLion did not reproduce \textbf{any} of these optimizations.
This observation suggests that \toolname can capture optimization opportunities beyond the reach of rule-based optimization tools.

\textbf{Python Version:} The maximum improvement on each Python project ranges from 61.77\% to 479.90\%, with an average improvement between 40.19\% and 143.25\%. 
Each Python project contains 4 to 11 test cases, among which 4 to 11 test cases exhibit improvements exceeding 10\%.

\textbf{Semantic Correctness Validation:} The authors further manually validated the optimized code for semantic correctness. All optimizations for both C/C++ and Python projects were semantically equivalent to the original programs, confirming the correctness of \toolname across both languages.

\begin{table}
\centering
\caption{Compile and test success rates.}
\label{tab:eval2-accuracy}
\vspace{-.5em}
\begin{adjustbox}{max width=1.1\linewidth}
\begin{tabular}{
    >{\centering\arraybackslash}m{1.2cm}
    >{\centering\arraybackslash}m{2.2cm}
    >{\centering\arraybackslash}m{1.5cm}
    >{\centering\arraybackslash}m{2.2cm}
    >{\centering\arraybackslash}m{2.2cm}
    >{\centering\arraybackslash}m{2.2cm}
}
\toprule
\textbf{Lang} & \textbf{Project} & \textbf{Functions} & \textbf{Total Opts} & \textbf{Compile Success} & \textbf{Test Success} \\
\midrule
\multirow{5}*{C/C++} & RocksDB     & 100 & 3585 & 2897 (80.81\%) & 2868 (80.00\%) \\
  & Redis       & 34  & 1200 & 1045 (87.08\%) & 1027 (85.58\%) \\
  & gRPC        & 30  & 1164 & 218 (18.73\%)  & 218 (18.73\%)  \\
  & LevelDB     & 47  & 1602 & 1366 (85.27\%) & 1318 (82.27\%) \\
  & spdlog      & 15  & 531  & 432 (81.36\%)  & 423 (79.66\%)  \\
\midrule
\multirow{5}*{Python} & click       & 55  & 3291 & - & 3291 (100.00\%) \\
  & flask       & 25  & 1587 & - & 1587 (100.00\%) \\
  & jinja      & 9   & 528  & - & 528 (100.00\%) \\
  & requests    & 46  & 3024 & - & 3024 (100.00\%) \\
  & scrapy      & 17  & 1053 & - & 1053 (100.00\%) \\
\bottomrule
\end{tabular}
\end{adjustbox}
\end{table}

\textbf{Compile and Test Success Rates:} As shown in Table~\ref{tab:eval2-accuracy}, for C/C++ projects, there are 226 hotspot functions, generating 8,082 optimization candidates, of which 73.72\% can be compiled successfully and 72.43\% can pass the tests. For Python projects, there are 152 hotspot functions, generating 9,483 optimization candidates, all of which pass the tests. These results indicate that the optimization candidates generated by \toolname have high compile and test success rates, ensuring practical utility in real-world applications without producing excessive invalid results.

The above results demonstrate that \toolname can effectively optimize real-world large-scale C/C++ and Python projects, significantly improving performance while maintaining semantic correctness, and the generated optimization candidates have high compile and test success rates, confirming its cross-language generalizability and strong practicality.

\textbf{Developer Feedback:} To further validate the practical value of \toolname, we submitted the optimizations generated by the C/C++ version to the corresponding open-source projects via GitHub Pull Requests. As of now, two projects (Redis and spdlog) have responded and accepted our optimization suggestions; the remaining three projects (RocksDB, LevelDB, and gRPC) have also received the Pull Requests, which are currently under code review, with LevelDB's Pull Request already receiving a positive acknowledgment from the community. This feedback demonstrates that \toolname has practical application value.

\subsection{A Complex Optimization Example Identified by \toolname and Accepted by Developers}\label{sec-5-example}

\begin{figure}[t]
    \centering
    \begin{subfigure}{\linewidth}
        \centering
        \includegraphics[width=.6\linewidth]{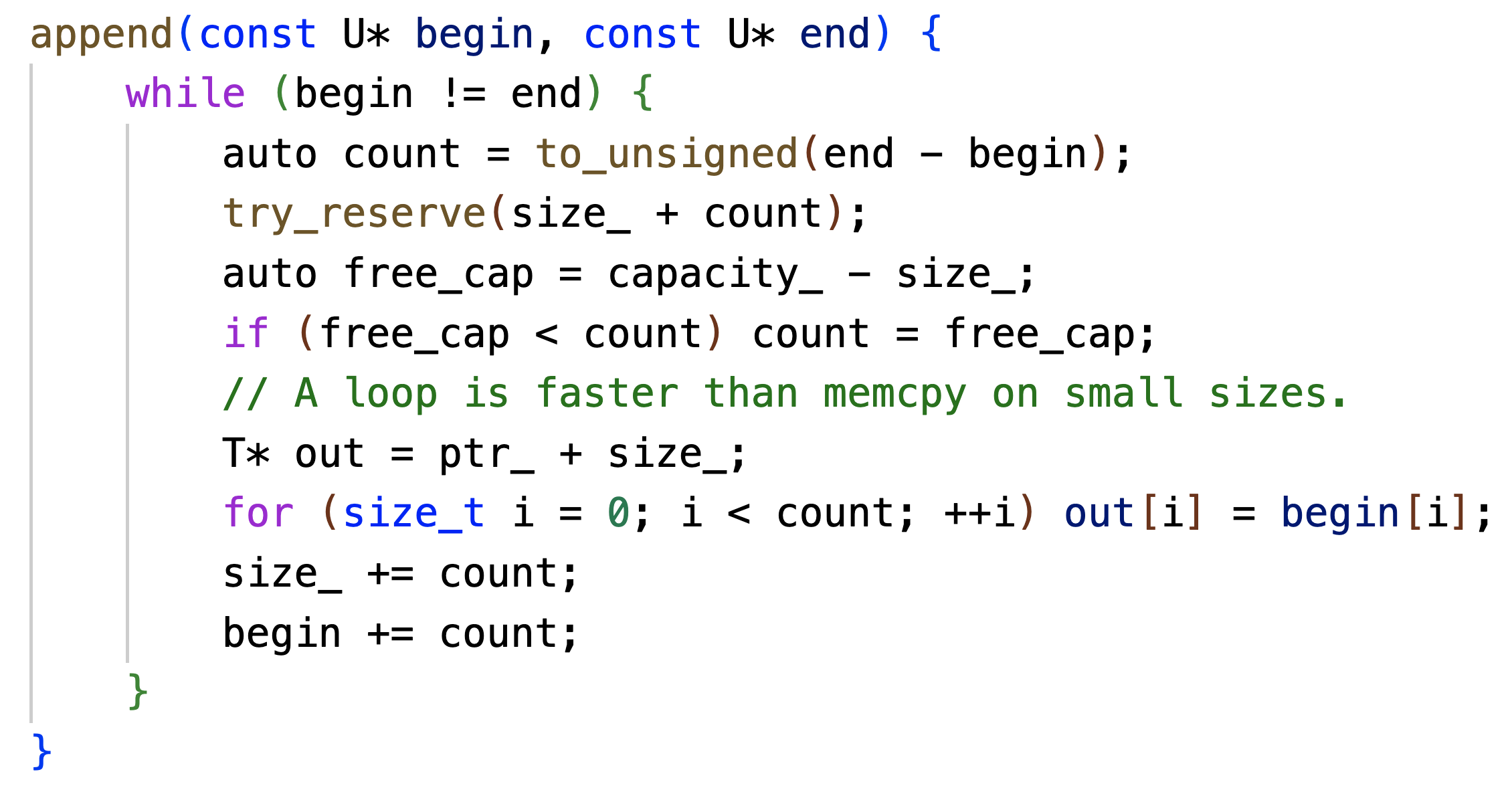}
        \caption{The code segment requiring optimization.}
        \label{fig:5-code_before}
        \vspace{1em}
    \end{subfigure}
    \begin{subfigure}{\linewidth}
        \centering
        \includegraphics[width=.6\linewidth]{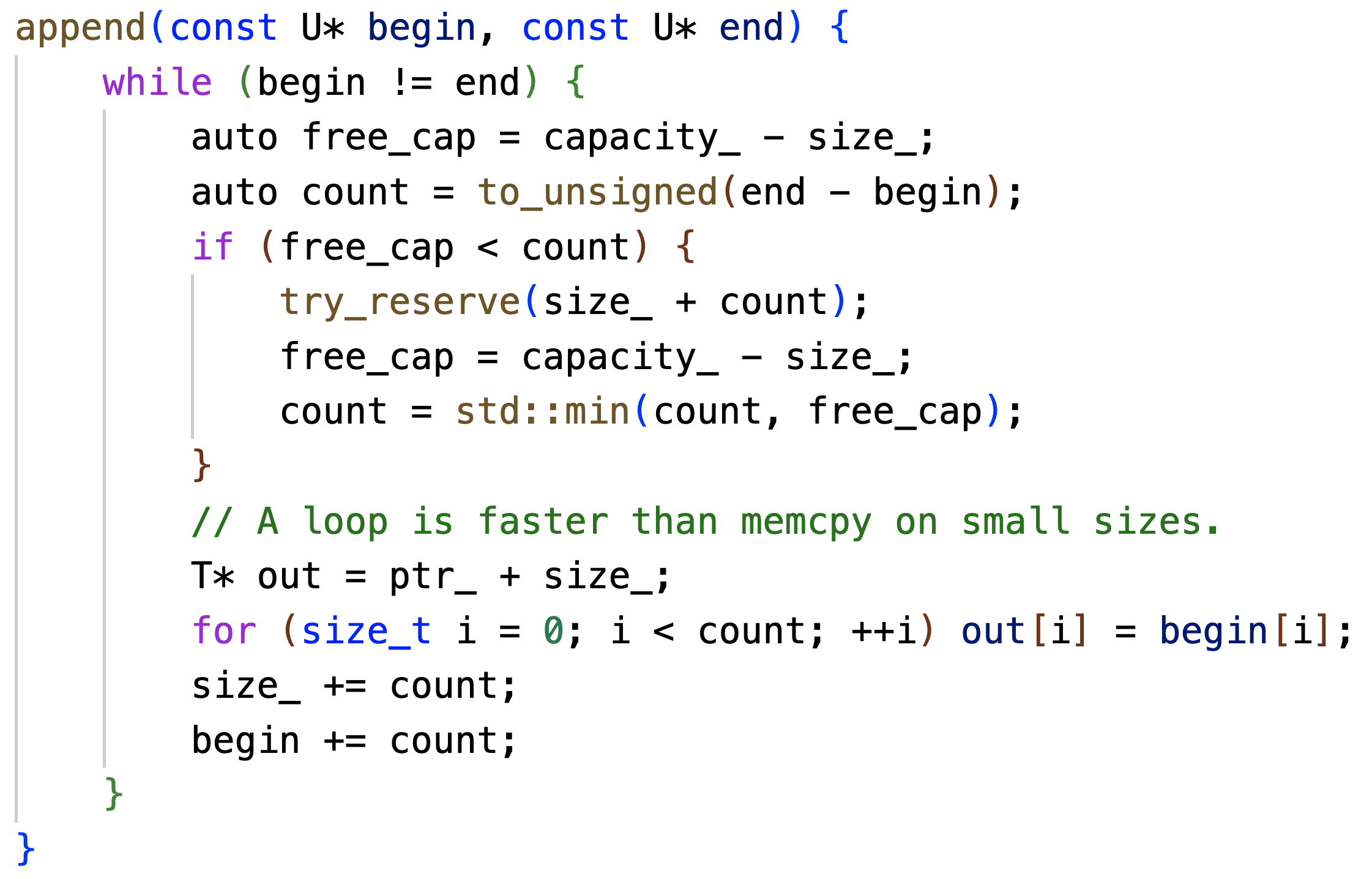}
        \caption{The optimization result.}
        \label{fig:5-code_after}
        \vspace{-.5em}
    \end{subfigure}
    \caption{A complex optimization identified by \toolname and accepted by developers.}
    \label{fig:5-example}
\end{figure}

To further demonstrate the capability of \toolname, Figure~\ref{fig:5-example} shows an optimization example generated in the previous evaluation. Notably, this optimization was accepted by the developers of the project, which demonstrates the correctness of the optimization.

The original implementation in Figure~\ref{fig:5-code_before} repeatedly invoked the \code{try\_reserve} function within a loop to ensure sufficient memory capacity.
\toolname correctly inferred the semantics of \code{try\_reserve}, which is defined in the \code{fmt} library as a non-standard memory pre-allocation routine. The optimization generated by \toolname in Figure~\ref{fig:5-code_after} reduces unnecessary calls to \code{try\_reserve} by first checking whether the current capacity is sufficient.

This change significantly reduces the number of memory allocation operations, improving performance, while preserving correctness. Importantly, this optimization requires understanding the project-specific effect of \code{try\_reserve}, which is not a part of the C++ standard library and whose semantics are not generally known to rule-based optimization tools. \toolname, by leveraging large language models, was able to infer the intent and usage of such library-specific constructs and suggest an effective optimization, thereby demonstrating its potential to handle less-general but impactful improvements in real-world projects.

\begin{rqbox}
\textbf{Finding:}
We demonstrate that \toolname can effectively optimize five large-scale C/C++ projects and five Python projects in the real world while maintaining correctness, confirming its cross-language applicability.
\end{rqbox}

\section{Discussion}\label{sec-discussion}

\subsection{Significance of Technological Contributions}

\toolname introduces two components as its main technical contributions: clustering optimization strategies, and generating corresponding Semgrep rules for each strategy. Omitting either component would result in a substantial increase in optimization overhead and a decline in optimization effectiveness:

\begin{itemize}
    \item \textbf{Clustering optimization strategies:} In the clustering results, we remove all noise points, and only 22\% of the commits are ultimately used to generate Semgrep rules. If this step were omitted, the overall optimization overhead would increase by 5x, and a substantial number of ineffective optimization strategies might be introduced, negatively impacting the final performance of \toolname.
    \item \textbf{Generating Semgrep rules:} Without utilizing Semgrep, given a piece of code requiring optimization, it would first be necessary to partition the code into multiple chunks and then use an LLM to evaluate, one by one, whether each strategy in the strategy library is applicable. In our C/C++ benchmark with 151 code optimization tasks, assuming each input code can be divided into 5 chunks and the strategy library contains approximately 150 optimization strategies, this would require $151 * 5 * 150 = 113,250$ LLM invocations in total. Such additional computational overhead is too costly for practical use.
\end{itemize}

\subsection{Correctness Guarantee}

\toolname employs multi-layer mechanisms to ensure the quality of optimization suggestions, including Semgrep rule location, LLM optimization, compilation verification, and test verification. These mechanisms can significantly reduce the manual review workload. As described in Section~\ref{sec-5-results}, 73.72\% of optimization suggestions pass compilation and 72.43\% pass tests, demonstrating the effectiveness of our automated filtering.

Nevertheless, optimization suggestions must ultimately undergo human review before adoption. Even after multi-layer automatic filtering, the possibility of LLMs generating semantically inequivalent optimization solutions cannot be completely eliminated. This is a common characteristic shared by all LLM-assisted code generation and modification tools, where the final adoption decision remains with the developers.

\subsection{Language Choice and Generalizability}

We have implemented \toolname on both C/C++ and Python and conducted extensive experimental validation. As demonstrated in Section~\ref{sec-empirical-eval}, \toolname achieves significant performance improvements on both the C/C++ benchmark (151 tasks) and the Python benchmark (150 tasks). As shown in Section~\ref{sec-5}, \toolname successfully optimizes five C/C++ projects and five Python projects in the wild. These results provide preliminary evidence that the approach generalizes across programming languages.

The overall workflow of \toolname is designed to be language-agnostic and can be reused for any language supported by Semgrep, such as Java, Rust, and others. To extend \toolname to another language, only minor code modifications would be needed to adapt to the new programming language. After that, the entire process can run automatically, with the main investment being the time required to execute the workflow. The most time-consuming part is Step 1 in Figure~\ref{fig:2-pipeline}, which involves building the strategy library. These tasks, such as downloading the necessary code repositories, fetching all commits, and performing the commit-level filtering, are relatively time-intensive.

In future work, we plan to further expand the range of programming languages supported by \toolname, such as JavaScript, TypeScript, Rust, etc., to validate its generalization capability across more programming languages and serve a broader user community.

\subsection{Limitation}
\toolname focuses exclusively on optimizing code within individual functions because we aim to tackle the complex problem of code optimization in a step-by-step manner. Future work may extend \toolname to support code optimization at a broader scope, such as multi-function modifications. 

Additionally, the types of optimization strategies we can currently implement are constrained by the expressiveness of the Semgrep static analyzer. Incorporating a better analysis tool may extend the capability of \toolname.

\subsection{Threats to Validity}

The main internal threat comes from potential data leakage. We already excluded exact matches in the strategy library, but there is still a possibility of data leakage through LLMs, since some experiment subjects may be included in their training set. However, as demonstrated in Section~\ref{sec-empirical-eval}, directly prompting LLMs to optimize code yields suboptimal results. \toolname outperforms other baselines with the same LLM (hence under the same degree of data leakage). Furthermore, Section~\ref{sec-5} shows that \toolname can generate new optimization suggestions for multiple projects in the wild, which are unlikely to be memorized by the LLM. These findings indicate that this threat has a limited impact on our findings.

One external threat comes from the selection of LLMs and the benchmark.
For LLMs, we evaluated \toolname on three popular LLMs from different vendors, all of which demonstrated that \toolname achieves superior optimization performance over baseline methods. Moreover, our implementation of \toolname is model-agnostic, and we provide the replication package so that other researchers can experiment on arbitrary LLMs.
For the benchmark, we selected the 100 most starred code repositories on GitHub for both C/C++ and Python separately, and randomly selected eligible commits from these repositories to serve as our benchmarks. 
Given the representativeness and diversity of the benchmark problems, we believe the external threat to the benchmark is limited. Furthermore, our approach has successfully optimized large-scale projects in the wild, further confirming the effectiveness of our approach.

\section{Related Work}\label{sec-related-work}

Early code optimization research mainly used rule-based and analytical methods targeting specific inefficiencies~\cite{mckeeman1965peephole,cocke1970global,della2015performance,krishna2020cadet,giavrimis2021genetic,clion,clangtidy}. These approaches rely on expert-defined rules, which are labor-intensive and difficult to scale. In contrast, \toolname automatically applies diverse optimization strategies and offers strong scalability.

With the popularity of deep learning, researchers have explored to train dedicated performance-improving models, such as VQ-VAE~\cite{chen2022learning}, Supersonic~\cite{chen2024supersonic}, and DeepDev-PERF~\cite{garg2022deepdev}. However, these dedicated models have been shown to have lower performance than LLM-based approaches~\cite{garg2023rapgen}, probably because the knowledge obtained from large-scale pre-training is often critical for performing the optimization. 

Recently, researchers also explored LLM-based approaches for code optimization. \citet{shypula2023learning} builds a benchmark for code optimization and evaluates the performance of various prompting techniques on this benchmark. A key finding of this study is that RAG could significantly outperform direct prompting. Our work confirms this finding and further proposes a new approach that significantly outperforms RAG. 
RAPGen~\cite{garg2023rapgen} performs API replacement-based code optimization for C\#. Compared to \toolname, RAPGen is limited to a single optimization strategy, whereas \toolname supports diverse strategies.
SBLLM~\cite{gao2024search} combines multiple optimized versions of the same code segment for better overall optimization. 
EffiCoder~\cite{huang2024swiftcoder} identifies performance bottlenecks by executing test cases and utilizes the obtained profiling data to guide LLMs in optimizing inefficient code segments, while PerfCodeGen~\cite{peng2025perfcodegen} employs bottleneck test cases as iterative feedback signals to prompt LLMs for progressive code refinement. 
Compared to our work, the latter three studies all explore orthogonal aspects and could potentially be combined with \toolname in the future.

Recent work has also explored using LLMs to synthesize static analysis checkers. KNighter~\cite{yang2025knighter} and AutoChecker~\cite{liu2024write} leverage LLMs to generate checkers for vulnerability detection, while MoCQ~\cite{li2025automated} adopts a neuro-symbolic approach for similar purposes. These works consistently demonstrate that synthesizing high-precision checkers is a non-trivial task, and the generated checkers often contain semantic errors that are difficult to fix. Unlike these approaches, \toolname shifts the correctness responsibility to the LLM optimization and downstream validation stages, allowing the rule generation stage to relax precision requirements moderately and focus more on recall, thereby alleviating the difficulties associated with directly synthesizing checkers to some extent.

\section{Conclusion}\label{sec-conclusion}

In this paper, we propose \toolname, an LLM-based automatic code optimization approach that integrates static program analysis techniques.
\toolname consists of three components: an optimization strategy library builder, an automatic static program analysis rule generator, and a library-based optimizer.
Extensive experiments conducted on three popular LLMs demonstrate that \toolname can effectively localize and retrieve appropriate optimization strategies on large projects, thereby guiding the LLMs to perform code optimizations. Furthermore, our in-the-wild evaluation on five C/C++ projects and five Python real-world projects shows that \toolname significantly improves the performance. These experiments together validate that the approach generalizes across programming languages.


\section*{Data Availability}

Artifacts of this paper, including the implementation of \toolname, all baseline implementations, and evaluation results, are available at \href{https://figshare.com/s/0391db49a3620ed53a1f}{https://\allowbreak{}figshare.com/\allowbreak{}s/\allowbreak{}0391db49a3620ed53a1f}.

\begin{acks}
  We thank all anonymous reviewers for their constructive feedback. We also thank Yujie Liu and Junqi Xu for their helpful discussions and support. This work is sponsored by the National Key Research and Development Program of China under Grant No. 2022YFB4501902, the National Natural Science Foundation of China under Grant No. 92582202, and Huawei Technologies Co., Ltd.
\end{acks}

\clearpage

\bibliographystyle{ACM-Reference-Format}
\bibliography{tosem-ref}

\section{Appendix}\label{sec-appendix}

This appendix presents all keywords and complete prompt templates used in the \toolname pipeline. Section~\ref{sec-9-1} lists the keywords used for searching performance-related commits on GitHub. Section~\ref{sec-9-2} presents the complete prompt templates used throughout the four stages of the pipeline: filtering optimization commits, summarizing optimization strategies, generating Semgrep rules, and generating code optimization results.

\subsection{Keywords Used in Section~\ref{sec-3-1-1}}\label{sec-9-1}

The data collection phase uses the following keywords when searching GitHub for performance-related commits (Section~\ref{sec-3-1-1}):
"performance", "optimize", "optimization", "speedup", "speed up", "speed-up", "fast", "faster", "fastest", "efficient", "efficiency", "throughput", "latency", "low-latency", "accelerate", "acceleration", "improve", "improvement", "enhance", "enhancement", "boost", "boosted", "tune", "tuning", "refactor for speed", "perf gain", "perf win", "reduce overhead", "reduce memory", "memory usage", "memory footprint", "memory consumption", "reduce allocation", "cache friendly", "cache efficiency", "cache hit", "cache miss", "inlining", "inline function", "loop unrolling", "vectorization", "vectorize", "simd", "parallelization", "parallelize", "multithreading", "multi-threading", "memoization", "lazy evaluation", "lock-free", "zero-copy", "hot path", "branch prediction", "prefetch", "bottleneck", "hotspot", "slow path", "critical path", "profile guided", "profiler", "complexity", "buffer reuse", "avoid copy", "avoid allocation", "reduce copy", and "batch processing".

\subsection{Prompt Templates}\label{sec-9-2}

\subsubsection{Optimization Commit Filtering Prompt}
\label{sec-9-2-1}

This prompt is used in the LLM-based commit filtering step of the data collection phase (Section~\ref{sec-3-1-1}) to determine whether a Git commit represents a genuine performance optimization. The prompt takes as input the commit message and the code diff, and outputs a binary true/false decision.

\bigskip
\noindent\textbf{System Prompt:}

\begin{verbatim}
You are an expert in software performance optimization and Git commit analysis.
Your task is to determine whether a given Git commit meets ALL of the following criteria:

1. The primary purpose of the commit is performance optimization
   (specifically reducing runtime resource consumption such as execution time
   or memory usage)
2. The optimization technique used is relatively generic and transferable
   to other functions or even other codebases
3. The changes do not primarily focus on code readability, maintainability,
   or other non-performance-related improvements

Performance optimization must be the MAIN goal, and the optimization approach
should be something that could potentially be applied elsewhere (e.g., algorithmic
improvements, data structure optimizations, caching strategies, loop
optimizations, etc.).

You must answer strictly with "true" or "false":
- Answer "true" ONLY if the commit meets ALL three criteria above
- Answer "false" if the commit does not meet any of the criteria or if the
  optimization is too specific/context-dependent

Do not provide any explanation, reasoning, or additional text in your response.
Only return "true" or "false".
\end{verbatim}

\bigskip
\noindent\textbf{User Prompt (fill-in template):}

\begin{verbatim}
Here is the information for a Git commit:

Commit Message:
{commit_message}

Git Diff:
{diff_content}

This commit only modifies one function in one file.
Based on the information provided above, does this commit meet all the following criteria:
1. Primary purpose is performance optimization (reducing runtime resource consumption)
2. Uses a relatively generic optimization technique that could be transferred
   to other functions/codebases
3. Not primarily focused on readability or maintainability improvements

Answer "true" only if ALL criteria are met, otherwise answer "false".
\end{verbatim}

\subsubsection{Optimization Strategy Summarization Prompt}
\label{sec-9-2-2}

This prompt is used in the strategy summarization phase (Section~\ref{sec-3-1-2}) to generate a one-sentence summary of the optimization strategy employed in a commit.

\bigskip
\noindent\textbf{System Prompt:}

\begin{verbatim}
You are a helpful assistant.
\end{verbatim}

\bigskip
\noindent\textbf{User Prompt (fill-in template):}

\begin{verbatim}
Please analyze the following commit that implements performance optimization.
Your task is to provide a ONE-SENTENCE summary in English describing an
optimization strategy that can be applied to similar code patterns.
The summary must be exactly one sentence - no more, no less.

IMPORTANT: If the optimization matches one of the common patterns below,
use the provided template summary. Otherwise, create a summary following
the same style.

COMMON OPTIMIZATION PATTERNS AND THEIR SUMMARIES:
1. Changing 'container.size() == 0' to 'container.empty()' -->
  Replace size() == 0 checks with empty() method calls for better
  performance and readability.
2. Reordering conditions in if statements with multiple 'and'-connected
   sub-conditions -->
  Reorder sub-conditions in if statements with multiple 'and'-connected
  conditions to place simpler or less expensive checks before more complex
  ones for better short-circuit evaluation.
3. Caching repeated calculations -->
  Cache frequently computed values to avoid redundant calculations in loops
  or repeated calls.

GUIDELINES for creating optimization summaries:
1. Include the main code structure or pattern where the optimization applies
   (e.g., 'in for loops', 'when using containers', 'in recursive functions')
2. Describe the specific change made within that code context
3. Include relevant technical details that are important for applying
   the optimization
4. Make the description specific enough to identify similar code patterns
   but general enough to apply across different projects

EXAMPLE of including code context:
- If a commit changes from pass-by-value to pass-by-reference in for loop
  iteration: Use pass-by-reference instead of pass-by-value when iterating
  over containers in for loops to avoid unnecessary object copying.
- If a commit optimizes string concatenation in loops: Use string builders
  or pre-allocated buffers instead of repeated string concatenation operations
  within loops to reduce memory allocations.
- If a commit changes vector push_back to reserve: Pre-allocate container
  capacity using reserve() before adding multiple elements to avoid repeated
  memory reallocations during insertion.

This commit has already been confirmed as a performance optimization.
Analyze the code changes and create a summary that includes both the code
context and the specific optimization technique.

Repository: {repo_name}
Commit Hash: {commit_hash}
Commit Message: {commit_message}
Modified Files: {modified_files}
Modified Functions: {modified_func}

Code Changes:
{diff_content}

Please provide your response as exactly one sentence that describes both the
code context (where to apply) and the optimization strategy (what to change).
\end{verbatim}

\subsubsection{Semgrep Rule Generation Prompts}
\label{sec-9-2-3}

These prompts are used in the Semgrep rule generation phase (Section~\ref{sec-3-2}) to convert an optimization strategy (represented as a code diff) into a Semgrep static analysis rule.

\bigskip
\noindent\textbf{System Prompt:}

\begin{verbatim}
You are an expert in C/C++ program analysis and Semgrep rule creation.
Your task is to identify performance optimization patterns in code changes
and create Semgrep rules that can detect similar optimization opportunities
in other codebases.

When writing Semgrep rules for C/C++, please ensure that:
1. The rule follows proper Semgrep YAML format with correct metadata
   including 'id', 'message', 'languages', 'severity', and 'pattern' fields.
2. The rule must support both C and C++ languages with 'languages: [cpp, c]'
   in the metadata.
3. The rule should detect patterns where code could be optimized similar
   to the provided diff.
4. The rule should focus on performance improvements rather than
   security issues.
5. Include helpful messages that explain the performance impact and
   suggested improvements.
6. Ensure that the generated YAML code is enclosed within triple backticks
   with 'yaml' language specification (```yaml).
7. The pattern should be syntactically correct for C/C++ and follow Semgrep
   pattern syntax.
8. Use appropriate Semgrep metavariables (like $VAR, $EXPR, $STMT) to make
   patterns generic.
9. Include severity level appropriate for performance optimizations
   (usually 'info' or 'warning').
\end{verbatim}

\bigskip
\noindent\textbf{Step 1 - Pattern Analysis (User Prompt):}

\begin{verbatim}
What performance optimization pattern does the following diff implement?
Analyze the changes and identify the optimization technique being applied.
Do not write Semgrep rules at this time, just analyze the diff and explain
the performance improvement.

Diff content:
{diff_content}
\end{verbatim}

\bigskip
\noindent\textbf{Step 2 - Rule Generation (User Prompt):}

\begin{verbatim}
Write a Semgrep rule for C/C++ that can detect similar optimization opportunities
in other codebases. The rule should identify code patterns that could benefit from
the same optimization technique shown in the diff. Focus on creating a rule that
helps developers find performance improvement opportunities. Make sure the rule
follows proper Semgrep YAML format, uses correct C++ pattern syntax, and supports
both C and C++ languages with 'languages: [cpp, c]'.
\end{verbatim}

\bigskip
\noindent\textbf{Step 3 - Error Fixing (User Prompt, triggered when Semgrep reports execution errors):}

\begin{verbatim}
There was an execution error when running the Semgrep rule:
Error: {error_message}

The YAML rule you provided:
{yaml_code}

Please fix the Semgrep rule to resolve this execution error.
Provide a complete, corrected YAML rule that can run successfully.
\end{verbatim}

\bigskip
\noindent\textbf{Step 4 - Semantic Fixing (User Prompt, triggered when Semgrep runs but misses the target code):}

\begin{verbatim}
The Semgrep rule execution completed but reported serious errors:

Full JSON result:
{json_result}

The YAML rule you provided:
{yaml_code}

Please analyze the errors and fix the Semgrep rule. Focus on:
1. Correct YAML syntax and structure
2. Valid C++ pattern syntax for Semgrep
3. Proper use of metavariables
4. Correct rule metadata

Provide a complete, corrected YAML rule that addresses all reported errors.
\end{verbatim}

The generation process iteratively applies Steps 3 and 4 until the Semgrep rule executes without fatal errors. The maximum number of iterations is configured as a hyperparameter (default: 3 rounds).

\subsubsection{Code Optimization Prompt}
\label{sec-9-2-4}

This prompt is used in the code optimization phase (Section~\ref{sec-3-3}) to generate optimized code for a given code segment, guided by the matched optimization strategy.

\bigskip
\noindent\textbf{User Prompt (fill-in template):}

\begin{verbatim}
Please optimize the following C/C++ function based on the identified code
segment and reference optimization strategies.

FUNCTION WITH LINE NUMBERS:
```cpp
{before_func_numbered}
```

TARGET CODE SEGMENT TO OPTIMIZE:
Lines {line_start} to {line_end}

REFERENCE OPTIMIZATION STRATEGIES:
Note: The following strategies may contain duplicates or overlapping suggestions.

{strategies_section}

ORIGINAL FUNCTION TO OPTIMIZE:
```cpp
{before_func}
```

REQUIREMENTS:
1. First, identify the specific code segment that needs optimization based on
   the given line numbers ({line_start}-{line_end})
2. Carefully analyze and understand each reference optimization strategy provided
3. Determine if each strategy is applicable to the identified target code segment
4. Verify that applying each strategy will preserve the original semantics
   and correctness
5. Only apply strategies that are both applicable and semantically safe
6. Maintain exact functionality throughout the entire function
7. Your response must contain exactly ONE code block with the complete
   optimized function. You may provide explanations and analysis in your
   response, but ensure there is ONLY ONE code block containing the entire
   optimized function.
\end{verbatim}

\bigskip
\noindent\textbf{Where the strategies section is constructed as follows:}

\begin{verbatim}
Strategy 1:
Optimization suggestion: {message_1}
Source: {repo_name}:{commit_hash_1}

Strategy 2:
Optimization suggestion: {message_2}
Source: {repo_name}:{commit_hash_2}

... (up to MAX_REFERENCE_STRATEGIES strategies, default: 4)
\end{verbatim}

The \texttt{strategies\_section} contains the top-$N$ optimization strategies (with $N$ = 4 by default) that Semgrep matched for the current code segment, each providing the strategy description and its source commit identifier. The complete optimization prompt for the motivating example in Section~\ref{sec-example} is illustrated in Figure~\ref{fig:2-prompt}.

\end{document}